\documentclass[final,12pt,authoryear,doi,times]{elsarticle}
\usepackage[utf8]{inputenc}
\usepackage{amsmath}
\usepackage{mathrsfs}
\usepackage{graphicx}
\usepackage{epstopdf}
\usepackage{enumitem}
\usepackage{siunitx}
\usepackage{float}
\usepackage{xcolor}
\usepackage{subcaption}
\usepackage{algorithm}
\usepackage{hyperref} %For clickable references
\usepackage{natbib}
    \makeatletter
\def\ps@pprintTitle{%
	\let\@oddhead\@empty
	\let\@evenhead\@empty
	\let\@oddfoot\@empty
	\let\@evenfoot\@oddfoot
}
\makeatother

\setlist{leftmargin=5cm}
\newcommand{\Rr}{\mathcal{R}^{\rm R}}

\newcommand{\Rs}{\partial \mathcal{R}_{\textbf{s}}^{\rm R}}

\newcommand{\Ri}{\partial \mathcal{R}_{\textbf{i}}^{\rm R}}

\newcommand{\vs}[1]{\textcolor{blue}{#1}}

\journal{Journal of the Mechanics and Physics of Solids}
 
\begin{document}

\title{Morphogenesis and proportionate growth: A finite element investigation of surface growth with coupled diffusion}

\author[ETH,cee]{Virginia von Streng}

\author[me]{Rami Abi-Akl}
\author[aa]{Bianca Giovanardi}
\author[me,cee]{Tal Cohen\corref{cor1}}
\ead{talco@mit.edu}

\address[ETH]{Department of Mechanical and Process Engineering, ETH Zurich, Switzerland}
\address[cee]{Department of Civil and Environmental Engineering, Massachusetts Institute of Technology, Cambridge, MA 02139}
\address[me]{Department of Mechanical Engineering, Massachusetts Institute of Technology, Cambridge, MA 02139}
\address[aa]{Department of Aeronautics and Astronautics, Massachusetts Institute of Technology, Cambridge, MA 02139}

\cortext[cor1]{Corresponding author.}

\begin{abstract}
Modeling the spontaneous evolution of morphology in natural systems and its preservation by proportionate growth remains a major scientific challenge. Yet, it is conceivable that if the basic mechanisms of growth and the coupled kinetic laws that orchestrate their function  are accounted for, a minimal theoretical model may exhibit similar growth behaviors. The ubiquity of surface growth, a mechanism by which material is added or removed on the boundaries of the body, has motivated the development of theoretical models, which can capture the diffusion-coupled kinetics that govern it. However, due to their complexity, application of these models has been limited to simplified geometries. In this paper, we tackle these complexities by developing a finite element framework to study the  diffusion-coupled growth and morphogenesis of finite bodies formed on uniform and flat substrates. We find that in this simplified growth setting, the evolving body exhibits a sequence of distinct growth stages that are reminiscent of natural systems, and appear spontaneously without any externally imposed regulation or coordination. The computational framework developed in this work can serve as the basis for future models that are able to account for growth in arbitrary geometrical settings, and can shed light on the basic physical laws that orchestrate growth and morphogenesis in the natural world.
\end{abstract}
\maketitle

\section{Introduction}
{\noindent From the life cycle of a living organism to that of an engineered artifact, the evolution of material systems is driven by competing processes of growth and decay. When these processes are coupled to simultaneously occurring adaptations, such as accumulation of mechanical stresses, deformation, diffusion, and chemical reactions, predicting the evolution of the material system becomes a nontrivial task. In the animal kingdom, these growth processes are often described as either morphogenesis \citep{turing1952chemical,cross1993pattern}, or proportionate growth \citep{sadhu2012modelling,bischofberger2016fluid}. The former refers to  changes in shape of an evolving body, such as the breakage of symmetry in an embryo\footnote{An embryo is initially spherical in its  blastula stage.}, and the latter to changes in size that preserve the overall shape, such as the growth of a child into an adult. From the biological viewpoint,  the orchestration of these complex growth processes is encoded in the DNA \citep{johnson1997molecular,metzger1999genetic}. Nonetheless, in recent years mechanics has emerged as a key player in driving natural growth \citep{savin2011growth,tallinen2016growth,budday2014}, thus potentially providing a link between the biological growth factors and the physical growth process, and in turn providing basic understanding that can guide the development of intelligent fabrication technologies that mimic natural growth.}

In this work, we focus our attention on material systems that can grow or degrade by reactions that occur on the boundaries of the body (i.e. surface growth) while accounting for the influence of diffusion through the bulk of the body that is needed to `feed' the growth, and the resulting development of mechanical stresses.   
Despite the ubiquity of their manifestations that range from examples of growth and morphogenesis in nature  \citep{Thompson} and to a multitude of engineering applications, these growth processes remain poorly understood. In nature, actin gel in eukaryotic cells grows by persistent polymerization that pushes against the cell membrane \citep{Noireaux,TCA}. This surface growth mechanism is also responsible for cell motility \citep{Mogilner96, Mogilner2003-Force}. Recent studies have shown that the rate of degradation of biopolymer particles in the oceans is mediated by mechanical stresses \citep{biopolymer, cordero2017}. Addition of layers underneath the bark of a tree and the growth of sea shells and horns serve as additional examples \citep{Skalak97,Skalak82, Moulton}.  In engineering, examples range from the formation of a solid electrolyte interface  that limits battery life \citep{horstmann2018review}, to drug delivery systems \citep{slaughter,li2016}, and emerging fabrication techniques induced by frontal polymerization \citep{robertson2018rapid, goli2019frontal}.  

In all of the examples mentioned above, the interaction between a solid backbone (e.g. a polymer network) and a diffusing fluid (e.g. a solvent) affects the rates of growth and degradation. The theoretical representation in the bulk must therefore account for both species (as in \cite{Hong,Chester,Duda,stracuzzi2018}), and the representation on the boundaries of the body should allow for chemical reactions between them, while obeying mass conservation requirements. Using this approach,  \cite{abiakl}  presented a general continuum framework for modeling the kinetics  of surface growth with coupled diffusion. Various alternative approaches to modeling surface growth have been proposed \citep{DiCarlo, Ciarletta,  Holland, Papastavrou, swain2018biological, Skalak82, Skalak97, Menzel, Moulton,TCA,ganghoffer2018, sozio2017,abeyaratne2020treadmilling, ateshian2007theory} and are reviewed in more detail by  \cite{abiakl}. 

Recently,  \cite{sozio2019nonlinear} presented a geometric theory that captures the nonlinear mechanics and incompatibilities induced by accretion, and studied various growth scenarios for given  vector fields of the growth velocity.  Using a different approach, \cite{zurlo2017printing,zurlo2018inelastic} study different methods for controlling the accretion process and the resulting non-Euclidean bodies that form. 
In contrast, the theoretical formulation in the present work does not prescribe the  kinematics of growth. Instead, it employs a kinetic law that relates between the growth velocity and its thermodynamic conjugate: the driving force of growth. The thermodynamically consistent form of the kinetic law is then chosen in accord with the specific mechanism of accretion. 

To study the chemo-mechanically coupled growth process,  \cite{abiakl} specialized the general model to a specific initial-boundary value problem of uniaxial growth. It was shown that although two separate time scales appear in the formulation (i.e. the diffusion time scale, and the growth time scale), the system rapidly adapts through a diffusion dominated response to arrive at a `universal path' that is independent of initial conditions; as it evolves along this path, growth and diffusion act harmoniously. 
The specific growth patterns considered in earlier work \citep{abiakl,abi2020surface} can be described by a single spatial coordinate and were thus amenable to both analytical and numerical investigation (using a finite difference scheme). To account for more complex geometries, in this paper we develop a finite-element model. We restrict our attention to 2D growth settings (i.e. generalized plane-strain) and take advantage of our \textit{a priori} knowledge on the existence of a `universal growth path', to reduce the complexity of the numerical simulation.

The developed model decouples the growth process into diffusion-deformation and growth steps. Every diffusion-deformation step is solved as a boundary value problem in quasi-equilibrium. Applying the growth velocities resulting from the diffusion-deformation solution to the material domain leads to a new material geometry, which then is remeshed. Subsequently, the next diffusion-deformation problem is solved on the new mesh. A simplified approach for adding material on a flat rigid growth surface is shown to ease  the computation by neglecting the development of residual stresses due to differential growth along the association surface. It is then confirmed that this effect is negligible in the considered growth settings.

The paper is organized as follows: in Section  \ref{sec:equations}, we define the problem setting and briefly recapitulate the governing equations  that were developed by \cite{abiakl}. Then, in Section \ref{sec:model}, we discuss the numerical solution of the highly nonlinear and coupled system of equations while  highlighting the procedure to capture the time evolution of the domain. The results of three representative simulations are presented in Section \ref{sec:sim}. Following validation of the model against analytical results for the 1D case, growth of more general 2D systems  with varying boundary conditions are portrayed, and the distinct stages of growth that emerge in these systems are discussed. Finally, we provide concluding remarks in Section \ref{sec:conclusion}.

\section{Problem setting and governing equations}\label{sec:equations}

Consider a rigid and flat  substrate submersed in a solvent. A chemical reaction that can occur on this substrate promotes the formation of new solid material. As accretion progresses, a mixed species continuum body is formed. 
We consider each species to be  separately incompressible and distinguish between the current configuration of the body $\mathcal{B}$, which occupies the region $\mathcal{R}(t)$ in the physical space, and its solvent free (i.e. dry) image $\mathcal{R}^{\rm R}(t)$ in the reference space\footnote{From hereon, quantities in the reference space are denoted by the superscript $\left(\cdot\right)^{\rm R}$.  }.  Each material point that is attached to the body in the reference configuration, denoted by $\textbf{X}$, sustains its location throughout the growth process, and its mapping to $\mathcal{R}(t)$ is denoted by  $\textbf{x}(\textbf{X},t)$. 

Previous studies have shown that, due to build up of residual stresses,  description of the natural reference configuration of the grown body may require a higher dimensional reference space \citep{abiakl,TCA}. In this work we bypass this complexity by focusing our attention on flat substrates and making a simplifying assumption on the growth process (as described in detail in Section \ref{growth}). It will be shown that for the considered geometric setting and constitutive model, the effects that are omitted due to this simplification are negligible. Hence, this approach can be extended in the future for arbitrary growth surfaces.

\begin{figure}[H]
  \centering
    \includegraphics[width = 0.58\textwidth]{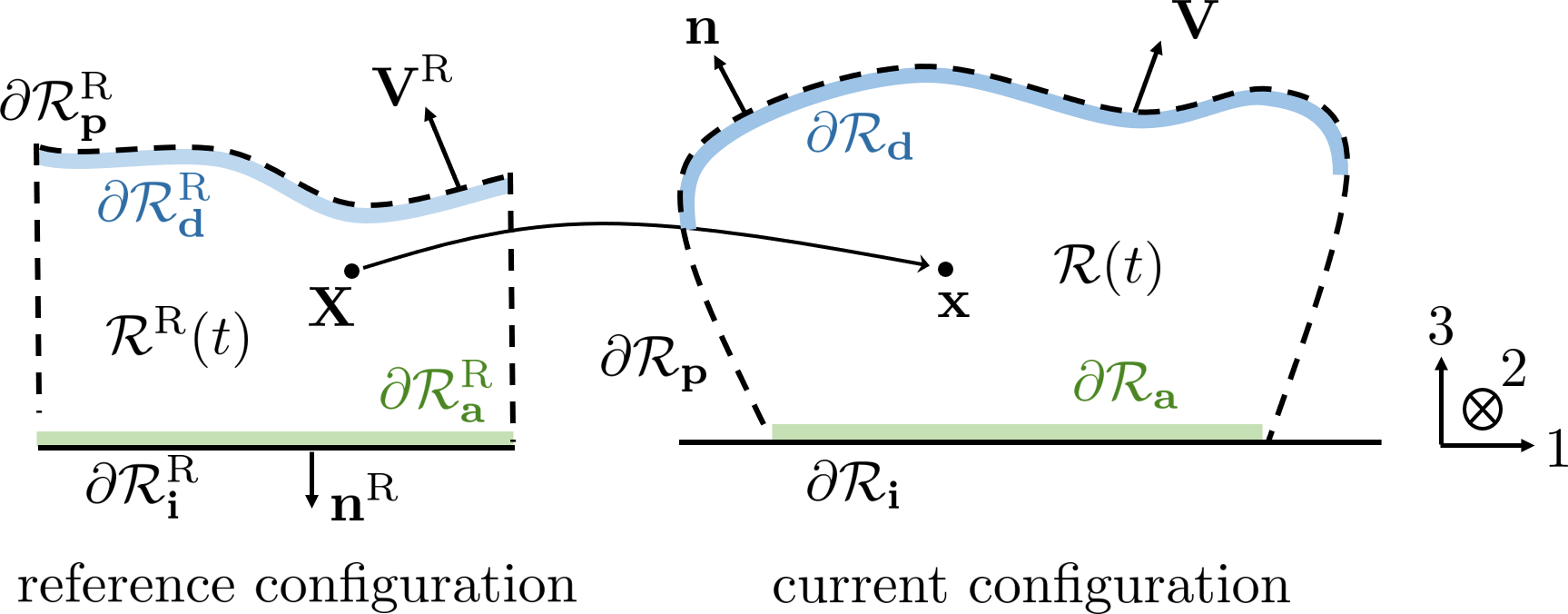}
  \caption{Illustration of the grown body in both current and reference configurations. Different regions of the boundary are shown by different style lines in both configurations. In this illustration, to represent the problem setting used to derive results in Section \ref{subsec:2D}, we set $\partial \mathcal{R}_{\textbf{p}} = \partial \mathcal{R}_{\textbf{s}}$ and $\partial \mathcal{R}_{\textbf{i}} = \partial \mathcal{R}_{\textbf{u}}= \partial \mathcal{R}_{\textbf{a}}$. However, this need not be the case. For example, the growth can occur only on a sub-region of $\partial\mathcal{R}_{\textbf{u}}$, as will be considered in Section \ref{subsec:1D}, or the substrate can be permeable.}
  \label{fig1}
\end{figure}

The deformation gradient is defined\footnote{The operator $\text{Grad}(\cdot)$ is the gradient with respect to the reference configuration, and $\text{grad}(\cdot)$ the gradient with respect to the current configuration. Similarly, the operator $\text{Div}(\cdot)$ is the divergence with respect to the reference configuration, and $\text{div}(\cdot)$ the divergence with respect to the current configuration. } as $\textbf{F}=\text{Grad}(\textbf{x})$ with the Jacobian $J=\text{det}(\textbf{F})$ representing the addition of volume due to solvent in the current configuration, and the displacement is given by $\textbf{u}(\textbf{X},t)=\textbf{x}(\textbf{X},t)-\textbf{X}$. We thus refer to $J$ as the swelling ratio and relate it to $\phi^{\rm R}$, the solvent volume fraction in the reference configuration, by the incompressibility constraint \begin{equation}\label{incompressibility}
    \phi^{\rm R}=J-1,
\end{equation} which assures that $J=1$ corresponds to the dry state.  The solvent volume fraction in the current state can be recovered from the transformation $\phi^{\rm R} = J \phi$.

The boundary of the  body $\mathcal{B}$ is denoted by $\partial \mathcal{R}(t)$ in the current and $\partial \mathcal{R}^{\rm R}(t)$ in the reference configuration. We classify the different (and potentially overlapping) regions on this boundary as follows:  $\partial \mathcal{R}_{\textbf{u}}(t)$ - is the association boundary where the material is fixed to the substrate (i.e. $\textbf{u}$ is constrained),  $\partial \mathcal{R}_{\textbf{s}}(t)$ - is the  unconstrained part  of the boundary, which may be subjected to external traction,  $\partial \mathcal{R}_{\textbf{d}}(t)$ - is the boundary on which material can dissociate, $\partial \mathcal{R}_{\textbf{a}}(t)$ - is the boundary on which material can associate, $\partial \mathcal{R}_{\textbf{i}}(t)$ - is the impermeable region of the boundary, and $\partial \mathcal{R}_{\textbf{p}}(t)$ - is the permeable region. Clearly, $\partial \mathcal{R}_{\textbf{u}}(t) \cap \partial \mathcal{R}_{\textbf{s}}(t) = \emptyset$ and 
$\partial \mathcal{R}_{\textbf{p}}(t) \cap \partial \mathcal{R}_{\textbf{i}}(t) = \emptyset$, while 
$\partial \mathcal{R}_{\textbf{d}}(t) \subset \partial \mathcal{R}_{\textbf{s}}(t)$ and $\partial \mathcal{R}_{\textbf{a}}(t) \subset \partial \mathcal{R}_{\textbf{u}}(t)$.
The complete boundary can be represented either by $\partial \mathcal{R}(t)= \partial \mathcal{R}_{\textbf{i}}(t)\cup \partial \mathcal{R}_{\textbf{p}}(t)$ or  $\partial \mathcal{R}(t)=\partial\mathcal{R}_{\textbf{u}}(t)\cup \partial\mathcal{R}_{\textbf{s}}(t)$, as illustrated in Fig. \ref{fig1}. The boundary of the body can move in both the  current and reference frames. If we denote a point on the boundary in the reference configuration by $\textbf{X}_b(t)$, then its image in the current configuration is $\textbf{x}_b(t)=\textbf{x}(\textbf{X}_b(t),t)$,  and the corresponding velocities of the boundary in the reference and current configurations are\begin{equation}\label{eq:boundary}
 \textbf{V}^{\rm R} =\frac{{\rm d} \textbf{X}_b}{{\rm d} t} \quad\text{and} \quad   \textbf{V} =\frac{{\rm d} \textbf{x}_b}{{\rm d} t},
\end{equation} respectively.  These velocities can be related by $\textbf{V}=\textbf{F}\textbf{V}^{\rm R}+\textbf{v}$ where $\textbf{v}=\partial \textbf{x}/\partial t$ is the material velocity.   In particular, we denote the  velocities of the association and dissociation boundaries by $\textbf{V}_\textbf{a}^{\rm R}$ and $\textbf{V}_\textbf{d}^{\rm R}$, respectively.

\subsection{Balance laws and boundary conditions}
In absence of body forces, momentum balance and corresponding boundary  conditions read
%We follow the formulation of Chester et al. \cite{Chester2015} to write  momentum and mass balance in the absence of body forces, to arrive at a  system of coupled partial differential equations with their respective boundary conditions: 
\begin{align}
{\rm Div} \left( \textbf{S} \right) = \textbf{0}\quad &\text{in} \quad \mathcal{R}^{\rm R}(t),\label{momentumbalance}\\
\label{u0_bc}
\textbf{u}={\bar{\textbf{u}}} \quad &\text{on} \quad {\partial \mathcal{R}^{\rm R}_{\textbf{u}}(t)},\\
\label{momentumneumann}
\textbf{S} \cdot \textbf{n}^{\rm R} = {\bar{ \textbf{s} } } \quad &\text{on} \quad{\partial \mathcal{R}_{\textbf{s} }^{\rm R}(t)},
\end{align}
where $\textbf{S}=\textbf{S}(\textbf{X},t)$ is the  first Piola–Kirchhoff stress tensor.  In \eqref{u0_bc} we prescribe the displacement constraint 
on the association surface, such that $\bar{ \textbf{u}}$ is determined by the growth process and the properties of the growth surface (as will be detailed in Section \ref{sec:model}), and in \eqref{momentumneumann} we write the stress free boundary condition on the free surface of the grown body, where $\textbf{n}^{\rm R}$ represents the outward facing normal to the boundary in the reference configuration (see Fig. \ref{fig1}).

In absence of an external source, mass balance is written as \citep{abiakl}
\begin{align}
\label{massbalance}
\dot{\phi}{^{\rm R}}+{\rm Div}(\textbf{j}^{\rm R})= \textbf{0} \quad &\text{in} \quad \mathcal{R}^{\rm R}(t),\\
\label{mu_cont}
\mu=\bar{\mu} \quad &\text{on} \quad \partial \mathcal{R}^{\rm R}_{\textbf{p}}(t),\\
\label{massneumann}
-\textbf{j}^{\rm R} \cdot \textbf{n}^{\rm R}= \bar{j}^{\rm R} \quad &\text{on} \quad {\partial \mathcal{R}^{\rm {R}}_{{\textbf i}}}(t),
\end{align}
where ${\textbf{j}^{\rm R}=\textbf{j}^{\rm R}(\textbf{X},t)}$ is the volumetric flux of solvent, $\mu$ is the chemical potential that drives the flux (as described in the next section), and  $\bar j^{\rm R}$ represents the penetrating flux through the inner side of the boundary. Although $\partial\mathcal{R}^{\rm {R}}_{{\textbf i}}$ is impermeable, the transformation of solvent species into solid species at this boundary creates a solvent flux that is determined by requiring conservation of volume across the boundary
\begin{equation}\label{growthbalance}
\bar{j}^{\rm R} = (J-1)\textbf{V}^{\rm R} \cdot \textbf{n}^{\rm R}.
\end{equation}
Here it is assumed that the solvent outside of the body is stationary. 

\subsection{Constitutive relations}
Following earlier studies \citep{Chester,abiakl,Hong, Duda} that employ  the Flory Rehner approach, the constitutive response of the two species material system can be described by its Helmholtz free energy per unit reference volume \citep{Flory42,Flory43} \begin{equation}\label{psi}
 \psi=\psi(\textbf{F},\phi^{\rm R}),   
\end{equation} which depends on the deformation of the solid network and concentration of solvent.  Constitutive relations for the Piola Kirchhoff stress and the chemical potential are readily derived using the Coleman-Noll procedure on the dissipation rate to write \begin{equation}\label{crf}
    \textbf{S}=\frac{\partial \psi}{\partial \textbf{F}}-pJ\textbf{F}^{-\rm T}\quad \text{and} \quad \mu=\frac{\partial \psi}{\partial \phi^{\rm R}}+p,
    \end{equation}
    respectively. Here, the hydrostatic pressure $p$ arises in response to the incompressibility constraint \eqref{incompressibility}. 
    
    The solvent flux can be related to the gradient of chemical potential via a  thermodynamically consistent kinetic law  that satisfies the non-negativity of the dissipation rate, i.e. $\textbf{j}^{\rm R}\cdot {\rm Grad}\mu\leq0$.
The specific response functions used in this work are detailed in Section \ref{resp_func}. 

Various forms of the coupled and nonlinear  systems of governing equations described above have been solved numerically by several authors, to model the swelling of fluid permeated soft solids \citep{wang2020high,liu2015multiplicative,papastavrou2013mechanics,liu2015dehydration,Hong,Chester2015,truster2017unified}, and different approaches have been proposed to  simultaneously capture the effects of large deformations,   incompressibility and swelling in bodies that are defined by a prescribed set of material points (i.e. with a constant $\mathcal{R}^{\rm R}$). 
An additional challenge in this work is imposed by the time evolution of the domain in both in the current and reference frames.  This evolution is determined by a growth law, which too is nonlinearly coupled to the deformation and diffusion response, as detailed next.

\subsection{Growth law}
By specializing the dissipation rate to find the contribution on the boundary of the body, and restricting growth to occur along the normal to the surface in the reference frame    a thermodynamically consistent law for the growth rate should obey the inequality \citep{abiakl} 
\begin{equation}\label{disip_ineq}
 \int\displaylimits_{\partial \mathcal{R}^{\rm R}} \left( \textbf{S}\textbf{n}^{\rm R} \cdot \textbf{F} \textbf{n}^{\rm R} + \Delta \psi + J\mu  \right) V^{\rm R} {\rm d}A\geq0,
\end{equation}
where $\textbf{V}^{\rm R}=V^{\rm R} \textbf{n}^{\rm R}$, and with the  \textit{latent energy of growth} defined as \begin{equation}\Delta\psi=\psi^+-\psi^-,\end{equation} where the superscripts $(\cdot)^+$ and $(\cdot)^{-}$ on a quantity denote its limiting values on the outer and inner side of the boundary, respectively. Since $V^{\rm R}$ is the rate of addition of solid mass (per unit area) we refer to it as the growth rate.   
From the above relation, it becomes apparent that 
the thermodynamic conjugate to this growth rate is \begin{equation}\label{driving_force}
    f = \textbf{S}\textbf{n}^{\rm R} \cdot \textbf{F} \textbf{n}^{\rm R} + \Delta \psi + J\mu, \end{equation}
which we refer to as the \textit{driving force of growth}. We require that  inequality \eqref{disip_ineq} is obeyed in any sub-region of the boundary surface, namely the growth can be described by a kinetic relation $V^{\rm R}=\mathcal{G}(f)$ for which $f\mathcal{G}(f) \geq0$.

\subsection{Formulation summary and specific constitutive response}\label{resp_func}

The formulation, written for an arbitrary set of constitutive response functions (i.e. the Helmholtz free energy $\psi(\textbf{F},\phi^{\rm R})$, the diffusion law, and the growth law $\mathcal{G}(f)$) provides a complete representation of the time dependent evolution of the two species growing body, by obeying requirements of momentum balance \eqref{momentumbalance} and mass conservation \eqref{massbalance},  and for  a given set of boundary conditions \eqref{u0_bc}, \eqref{momentumneumann} and \eqref{mu_cont}, \eqref{massneumann}.  The specific forms of the constitutive response functions used to derive the numerical results in this work are detailed next.   To compare the results of the present work with analytical results from the earlier study by \cite{abiakl}, we apply the same constitutive response functions that were used therein. 

For the response in the bulk, we employ the constitutive response functions which have been proposed for modeling large deformations in fluid permeated polymer networks \citep{Chester,Duda,Hong}. 
Accordingly, we decompose the free energy 
\eqref{psi} into an elastic energy $\psi_e$ due to deformation of the solid network and a mixing energy  $\psi_s$, such that
\begin{equation}\label{energy}
\psi(\textbf{F},\phi^{\rm R})=\psi_e(\textbf{F})+\psi_s(\phi^{\rm R}).
\end{equation}
The elastic energy is taken as \citep{Flory43,Flory42} 
\begin{equation}\label{elasticenergy}
\psi_e(\textbf{F})=\frac{G}{2} \left[ |\textbf{F}|^2-3-2 \ln  \left( \det \textbf{F} \right) \right], \quad G=nkT,
\end{equation}
where $G$ is the shear modulus, $n$ is the number of polymer chains per unit volume, $k$ is the Boltzmann's constant, and $T$ is the  temperature. The energy of mixing is taken as 
\begin{equation}\label{solventenergy}
\psi_s(\phi^{\rm R})=\phi^{\rm R} \left\{ \psi_0+\frac{kT}{\nu} \left[ \ln \left( 1-\frac{1}{1+\phi^{\rm R}} \right) +\frac{\chi}{1+\phi^{\rm R}} \right] \right\},
\end{equation}
where $\chi$ is the Flory Huggins interaction parameter, $\nu$ is the volume of a solution unit, and $\psi_0$ is the reference free energy of the solution. On the dissociation boundary (i.e. $\partial\mathcal{R}_{\textbf{d}}$) we can thus write $\psi^+=J\psi_0$ where multiplication by the volume ratio $(J)$ assures that the free energy is written per the same unit volume on either side of the boundary. On the association boundary (i.e. $\partial\mathcal{R}_{\textbf{a}}$), a chemical binding potential $\psi_a$ can  locally alter the potential energy and represents the energetic gain due to association, hence (as in \cite{abiakl}) we take $\psi^+=\psi_a$.

Substituting   \eqref{energy} with \eqref{elasticenergy} and \eqref{solventenergy} into the constitutive relations \eqref{crf}, yields the constitutive response functions 
\begin{equation}\label{eq:stresslaw}
\textbf{S} = G \left( \textbf{F} - \textbf{F}^{-T} \right) - Jp \textbf{F}^{-T},
\end{equation}
and
\begin{equation}\label{chempot}
\mu = \mu_0 + \frac{kT}{\nu}\left[\ln\left(1-\frac{1}{J}\right)+\frac{1}{J}+\frac{\chi}{J^2}\right] + p,
\end{equation}
where $\mu_0$ is the chemical potential of the surrounding fluid region. It is assumed to be constant, and in view of the incompressibility constraint is equal to the reference free energy (i.e. $\mu_0=\psi_0$). The above equation, allows us to write the pressure $p$ in terms of the chemical potential in the useful form
\begin{equation}\label{eq:pressure}
p = \mu - \mu_0 - \frac{kT}{\nu}\left[\ln\left(1-\frac{1}{J}\right)+\frac{1}{J}+\frac{\chi}{J^2}\right].
\end{equation}

A kinetic law for the solvent flux is chosen to obey the classical form \cite{Feynman}, which in the reference frame reads
\begin{equation}\label{fluxphir}
\textbf{j}^{\rm R} = -\frac{D \nu}{kT} (J-1) \textbf{F}^{-1} \textbf{F}^{-\rm T}  {\rm Grad} \mu,
\end{equation} where $D$ is the diffusion coefficient. This diffusion law further implies continuity of chemical potential across the permeable boundary. Hence, $\bar \mu=\mu_0=\mu^{+}=\mu^{-}$ on  $\partial\mathcal{R}_{\textbf{p}}$. 

The kinetic law for growth is chosen to  relate  the rate of polymerization $V^{\rm R}$ to the driving force $f$ of equation \eqref{driving_force} using the classical Arrhenius form. Both  association and dissociation reactions are considered in view of the dissipation inequality \eqref{disip_ineq},  by including two separate exponential terms, to write 
\begin{equation}\label{growthvelocity}
V^{\rm R} =\mathcal{G}( f ) = \frac{b}{2} \left( \rm{e} ^{\frac{\nu f}{kT}} - \rm{e} ^{-\frac{\nu f}{kT}} \right) = b \sinh \left(\frac{\nu f}{kT} \right) ,
\end{equation} where $b>0$ is the reaction constant. 

In summary, the constitutive response depends on a set of seven material parameters: $G,\nu$, $\chi$, $\psi_0$, $\psi_a$, $D$, $b$
and assuming room temperature we have $kT=\SI{4d-21}{J}$. For the  derivation of results in this work, we use values in the common range \citep{Chester,Hong,lai} for parameters that describe the bulk response  $G=\SI{4}{kPa}$, $\nu=\SI{d-28}{m^3}$, $\chi=0.2$, $\psi_0=\SI{-0.4}{MPa}$, $D=\SI{d-4}{m^2.s^{-1}}$. Parameters that describe the growth kinetics are taken as $\psi_a=\SI{-13}{MPa}$, and $b=\SI{d-7}{m.s^{-1}}$. The sensitivity analysis in this work will center on varying the boundary conditions to examine the effect of geometry on the growth process.  

\section{Numerical implementation}\label{sec:model}

In addition to the geometric nonlinearities arising from the very large 
deformations, the governing equations \eqref{momentumbalance}-\eqref{momentumneumann} and  \eqref{massbalance}-\eqref{massneumann} unveil several nonlinearities in the coupling mechanisms between the deformation and the chemical potential. On the one hand, an increase (decrease) of the chemical potential contributes to pressure build-up (drop) in the bulk material. On the other hand, the deformation of the bulk affects the diffusivity through the bulk via the chemical potential. Moreover, both the deformation and the chemical potential contribute to the driving force of growth  \eqref{driving_force}, which dictates the velocity of the boundary and constitutes an additional source of nonlinearity. In this section we propose a numerical scheme that addresses the challenges emerging in the simulation of this complex response.

\subsection{Weak form of the governing equations}\label{wf}
Equations \eqref{momentumbalance}-\eqref{momentumneumann} and  \eqref{massbalance}-\eqref{massneumann} represent two, coupled, boundary value problems for the primal unknowns $\textbf{u}$ and $\mu$. 
We introduce the test functions $\textbf{w}$ and $q$ for the displacement and for the chemical potential, respectively,  which vanish on $\partial \mathcal{R}_{\mathbf{u}}$ and $\partial \mathcal{R}_{\textbf{p}}$, respectively, and obtain the following weak formulations of the boundary value problems
\begin{equation}\label{wu}
\int\displaylimits_{\Rr(t)}\textbf{S}(\textbf{u}, \ \mu)  : {\rm Grad}\  \textbf{w} \ {\rm d}v = \int\displaylimits_{\Rs(t)} {\bar{ \textbf{s}} }\cdot \textbf{w} \ {\rm d}a \quad \forall \textbf{w},
\end{equation}
and
 \begin{equation}\label{ws}
 \int\displaylimits_{\Rr(t)} \textbf{j} ^{\rm R}(\textbf{u}, \ \mu) \cdot {\rm Grad} \ q \ {\rm d}v 
 -\int\displaylimits_{\Rr(t)} \dot{J} q \ {\rm d}v - \int\displaylimits_{\Ri(t)} (J-1) \ V^{\rm R} \ q \ {\rm d}a = 0\quad \forall q,
\end{equation}
where we have used  \eqref{growthbalance}, to express  the conservation of volume at the boundary, and  \eqref{incompressibility}, to express the material incompressibility. In the above integral representation, we have highlighted the dependence of the stress tensor $\textbf{S}$ and the flux $\textbf{j}^{\rm R}$ on the primal variables $\textbf{u}$ and $\mu$. We recall that $\textbf{S}(\textbf{u}, \ \mu)$ and $\textbf{j}^{\rm R}(\textbf{u}, \ \mu)$ are given by \eqref{eq:stresslaw}-\eqref{eq:pressure}, which we rewrite in the form
$$\textbf{S}(\textbf{u}, \ \mu) = G \left( \textbf{F} - \textbf{F}^{-T} \right) - J \left(\mu - \mu_0 - \frac{kT}{\nu}\left[\ln\left(1-\frac{1}{J}\right)+\frac{1}{J}+\frac{\chi}{J^2}\right] \right) \textbf{F}^{-T},
$$
and 
$$
\textbf{j}^{\rm R}(\textbf{u},\ \mu) = -\frac{D \nu}{kT} (J-1) \textbf{F}^{-1} \textbf{F}^{-\rm T}  {\rm Grad} \mu,
$$
where $\textbf{F} =\text{Grad}(\textbf{u})-\textbf{I}$ and $J=\text{det}\left(\text{Grad}(\textbf{u})-\textbf{I}\right)$.

\subsection{Time discretization}\label{subsec:time-semi-discretization}

We propose an iterative algorithm to advance the solution from time $t$ to time $t+\Delta t$. We employ a first-order discretization of the time derivatives to approximate $\dot{J}$ at time $t + \Delta t$ as
$$\dot{J}(t + \Delta t) \approx \frac{J(t + \Delta t) - J(t)}{\Delta t},$$
and the boundary velocity $\textbf{V}^{\rm R}$ in  \eqref{eq:boundary} at time $t + \Delta t$ as
$$
\textbf{V}^{\rm R}(t + \Delta t)=\frac{{\rm d} \textbf{X}_b}{{\rm d} t}(t + \Delta t) \approx \frac{\textbf{X}_b(t + \Delta t) - \textbf{X}_b(t)}{\Delta t}.
$$

Since the time scale of diffusion is significantly faster than that of growth, we treat equations \eqref{wu} and \eqref{ws} in a fully-coupled fashion, while decoupling them from the time evolution of the domain $\Rr(t)$. We have verified that this approach and the results shown in Section \ref{sec:sim} are at convergence with respect to the time step, which confirms that for sufficiently small time steps the evolution of the domain $\Rr(t)$ can be treated explicitly without loss of accuracy.

At time $t + \Delta t$, we seek the displacement field $\textbf{u}(t + \Delta t)$ and the chemical potential field $\mu(t + \Delta t)$ which satisfy the following set of equations

\begin{equation}\label{eq:semi-discrete:wu}
\int\displaylimits_{\Rr(t)}\textbf{S}\left(\textbf{u}(t + \Delta t), \mu(t + \Delta t)\right)  : {\rm Grad} \ \textbf{w} \ {\rm d}v = \int\displaylimits_{\Rs(t)} {\bar{ \textbf{s}} }\cdot \textbf{w} \ {\rm d}a \quad \forall \textbf{w},
\end{equation}
 \begin{equation}\label{eq:semi-discrete:ws}
 \begin{aligned}
\int\displaylimits_{\Rr(t)} \textbf{j} ^{\rm R}\left(\textbf{u}(t + \Delta t), \mu(t + \Delta t)\right) \cdot {\rm Grad} \ q \ {\rm d}v 
&-\int\displaylimits_{\Rr(t)} \frac{J(t + \Delta t) - J(t)}{\Delta t} q \ {\rm d}v  \\
&- \int\displaylimits_{\Ri(t)} (J(t+\Delta t)-1) \ V^{\rm R} (t + \Delta t) \ q \ {\rm d}a = 0 \quad \forall q.
\end{aligned}
\end{equation}

To solve the problem in a fully-coupled fashion we adopt a Newton-Raphson iterative algorithm  to advance from one nonlinear iteration to the next until convergence, starting from initial
guesses for $\textbf{u}$ and $\mu$ given by the converged solutions at time $t$. 

Upon convergence, we update the domain $\Rr(t)$ by updating its boundary with 
\begin{equation}\label{eq:semi-discrete:boundary}
\textbf{X}_b(t+\Delta t) = \textbf{X}_b(t) + \Delta t \ V^{\rm R}(t + \Delta t) \textbf{n}^{\rm R} (t),
\end{equation}
where the velocity $V^{\rm R}(t + \Delta t)$ is computed in terms of the newly updated
displacement $\textbf{u}(t + \Delta t)$ and chemical potential $\mu(t + \Delta t)$.
The new domain is then remeshed, as described in the next section. The algorithm to update the displacement, chemical potential and problem domain from time $t$ to time $t + \Delta t$ is further summarized in Algorithm \ref{alg:decoupling}.

\subsection{Space discretization and growth}\label{growth}

Our numerical simulations consider the growth of a body with an initially rectangular cross-section of dimensions $w^{\rm R}\times l^{\rm R}$ (in the reference configuration), and  assumes generalized plane-strain conditions, such that the motion occurs only in the plane of the cross-section, i.e. in the $(X_1,X_3)$ plane.  The association surface spans the width $w^{\rm R}$ along the  $X_1$ direction  and the initial length, along $X_3$, is $l^{\rm R}$. We take advantage of the symmetry of the problem to resolve only half of the domain\footnote{We have confirmed that solution over the entire domain leads to identical results.}. 

We perform the space discretization by introducing a triangular mesh of the computational domain $\Rr$. The dimensions of the domain in which growth occurs, i.e. $w^{\rm R}$ and $l^{\rm R}$, are each divided into $10$ elements across with further refinement near the edges. The dimension along $X_2$ is divided into 2 elements and appropriate boundary conditions with no displacements or flux applied to maintain generalized plane-strain conditions. First order shape functions are used for the chemical potential $\mu$ and second order shape functions for the displacement $\textbf{u}$.

\begin{figure}[H]
  \centering
    \includegraphics[width = 0.7\textwidth]{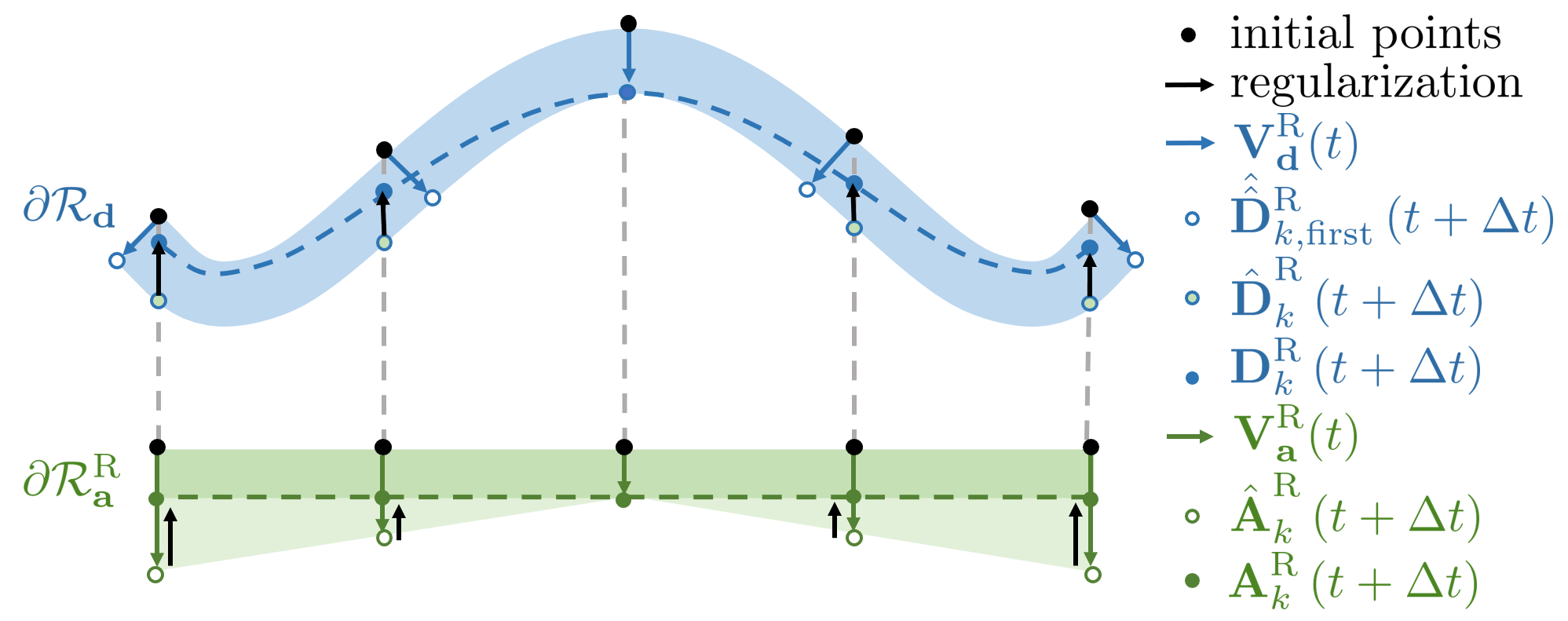}
  \caption{Schematic showing the construction of the boundaries of the body in the next time step and the volume preserving regularization. }
  \label{fig3}
\end{figure}

As discussed in Section \ref{subsec:time-semi-discretization}, we solve the fully-coupled set of equations for the displacement and chemical potential with a  
Newton-Raphson nonlinear solver. We solve the linearized system with a direct solver 
for the sake of robustness. More precisely, because the fully-coupled linear system is
not symmetric we use a LU-factorization solver for the linear steps.
Upon convergence of the nonlinear solver we evaluate the growth velocity at the boundary and accommodate growth by redefining the domain and creating a new mesh at every time step, based on the solution at the previous time. 
Note that this requires to reinterpolate the newly computed swelling ratio $J(t+\Delta t)$ on the new computational domain $\Rr(t + \Delta t)$ so that the second integral in Equation \eqref{eq:semi-discrete:ws} can be computed. 
Since the finite element mesh is three dimensional while in the considered generalized plane-strain deformation pattern motion is only permitted in the plane, the growth is reduced to two dimensions. Namely, only a grown/dissociated surface of incremental thickness is defined at every time step and then extruded in its normal direction following a volume conserving regularization procedure, as detailed next and illustrated on Fig. \ref{fig3}. 

At every time step, the association boundary in the reference configuration is defined by $N$ points, $\textbf{A}_1^{\rm R}(t), ..., \textbf{A}_N^{\rm R}(t)$, and the dissociation boundary in the reference configuration by $N$ points, $\textbf{D}_1^{\rm R}(t), ..., \textbf{D}_N^{\rm R}(t)$. Since motion is not permitted along $X_2$, the values of ${A}_{k,2}^{\rm R}$ and  ${D}_{k,2}^{\rm R}$ are constant in time.  

The association boundary of the new domain is determined via two steps. The intermediate step is 
\begin{equation}\label{eq:newnodes}
\hat{\textbf{A}}_{k}^{\rm R}\left(t+\Delta t \right) = \textbf{A}_k^{\rm R}\left(t\right) + {V}_{\textbf{a},k}^{\rm R}\left(t\right) \cdot \Delta t \cdot \textbf{n}_{{\textbf a},k}^{\rm R}, \qquad \forall k=1, ..., N,
\end{equation}
where the superimposed $\hat{(\cdot)}$ denotes the fact that this is an intermediate value, and with the index $k$ going through all $N$ nodes that lie on $\partial \mathcal{R}_{\textbf{a}}^{\rm R}$. Given the fixed boundary condition on the association surface, the normal vectors are constant and given by
\begin{equation}
\textbf{n}_{{\textbf u},k}^{\rm R}= \left( 0,0, -1 \right)^T, \qquad \forall  k=1, ..., N.
\end{equation}
The dissociation boundary of the new domain is determined via three steps. The first intermediate step is 
\begin{equation}\label{eq:newnodesD}
\hat{\hat{\textbf{D}}}_{k}^{\rm R}\left(t+\Delta t \right) = \textbf{D}_{k}^{\rm R}\left(t\right) + {V}_{\textbf{d},k}^{\rm R}\left(t\right) \cdot \Delta t \cdot \textbf{n}_{{\textbf d},k}^{\rm R}(t), \qquad \forall k=1, ..., N ,
\end{equation}
with the index $k$ going through all $N$ nodes that lie on $\partial \mathcal{R}_{\textbf{d}}^{\rm R}$. Unlike the association surface, the morphology of the dissociation surface changes in time and the normal vectors $\textbf{n}_{{\textbf d},k}^{\rm R}$ are calculated newly at every time step. The unnormalized vectors are defined as the outward pointing vector perpendicular to the connecting edge between the next and the previous node 
\begin{equation}
\hat{\textbf{n}}_{{\textbf d},k}^{\rm R}(t)=\begin{cases}
    \textbf{R} \cdot \left( \textbf{D}_{k+1}^{\rm R}(t)-\textbf{D}_k^{\rm R}(t) \right), & \text{if $k = 1$},\\ 
    \textbf{R} \cdot \left( \textbf{D}_{k+1}^{\rm R}(t)-\textbf{D}_{k-1}^{\rm R}(t) \right), & \text{if $k=2, ..., N-1$},\\
    \textbf{R} \cdot \left(\textbf{D}_k^{\rm R}(t)-\textbf{D}_{k-1}^{\rm R}(t) \right), & \text{if $k = N$},\\
  \end{cases}
\end{equation}with the rotation matrix 
\begin{equation}
\textbf{R} = \left[ \begin{matrix}
0 & 0 & 1\\
0 & 1 & 0 \\
-1 & 0 & 0
\end{matrix} \right],
\end{equation}
and are then normalized to yield
\begin{equation}
\textbf{n}_{{\textbf d},k}^{\rm R}(t) = \frac{\hat{\textbf{n}}_{{\textbf d},k}^{\rm R}(t)}{| \hat{\textbf{n}}_{{\textbf d},k}^{\rm R}(t) |}.
\end{equation}

Depending on the direction of the normal vectors, it is now possible that some points obtained from the first intermediate step \eqref{eq:newnodesD} have $\hat{\hat{D}}_{k,1}^{\rm R}\left(t+\Delta t \right)$ outside the  body in the reference configuration (Fig. \ref{fig3}). To avoid this, without loss of generality, we define the points to align with the  initially defined locations in the $X_1$ direction, such that $D_{k,1}^{\rm R}\left(t\right)=D_{k,1}^{\rm R}\left(0\right)$. To this end, the function ${\hat{D}}_{k,3}^{\rm R}\left(t+\Delta t \right) = f\left(\hat{\hat{D}}_{k,3}^{\rm R}\left(t+\Delta t \right)\right)$ is linearly inter- and extrapolated  to obtain $\hat{\textbf{D}}_{k}^{\rm R}(t+\Delta t)$.

Since the association boundary is subjected to a fixed boundary condition, its shape in the current configuration is preserved over time. Its shape in the reference configuration, however, is only preserved over time if the  growth rate  (i.e. $V^{\rm R}_{\textbf{a},k}\left(t\right) \cdot \Delta t \cdot \textbf{n}_{\textbf{a},k}^{\rm R}(t)$) is uniform  for all points k on the association boundary. If growth is not uniform, morphing of the association boundary in the reference configuration leads to residual stresses. Yet, to reduce the complexity in the present numerical model, we constrain the system to produce only uniform growth. We implement this assumption by including a volume preserving regularization in determining the reference location of the boundaries in the next time step
\begin{equation}
D_{k,3}^{\rm R}\left(t+\Delta t \right) =\hat{ D}_{k,3}^{\rm R}\left(t+\Delta t \right) + \hat{A}_{k,3}^{\rm R}\left(t+\Delta t \right) - \hat{A}_{0,3}^{\rm R}\left(t+\Delta t \right), \qquad \forall k=1, ..., N,
\end{equation}
\begin{equation}
A_{k,3}^{\rm R}\left(t+\Delta t \right) = \hat{A}_{0,3}^{\rm R}\left(t+\Delta t \right), \qquad \forall k=1, ..., N,
\end{equation}
where $\textbf{A}_0^{\rm R}$ is the point where the growth is minimal.

Although this regularization omits the physical effects that may arise in response to nonuniform growth. This assumption is exact for a chemically uniform  association surface, at both the limits of an infinitely long substrate $w/l\to\infty$ (i.e. uniaxial growth) and a very thin substrate $w/l\to0$ (i.e. filament growth). For the present simulations, it will be shown, by post evaluation (in Section \ref{regular}), that this effect is negligible in all of the considered settings. Nonetheless, future work will expand on the present framework to account for these effects, which may become significant if the chemical properties of the growth surface are nonuniform. 

\begin{algorithm}
Given $J(t)$ and $\mathcal{R}^{\rm R}(t)$, do:
\begin{enumerate}
\item Compute the updated $\mathbf{u}(t + \Delta t)$ and $\mathbf{\mu}(t + \Delta t)$ by solving the fully-coupled nonlinear problem \eqref{eq:semi-discrete:wu} and \eqref{eq:semi-discrete:ws} on $\mathcal{R}^{\rm R}(t)$;
\item Update $\mathbf{F}(t + \Delta t)$, $\mathbf{S}(t + \Delta t)$, $J(t + \Delta t)$ based on $\mathbf{u}(t + \Delta t)$;
\item Compute the updated boundary velocity $V^{\rm R}(t + \Delta t)$ from Equation \eqref{growthvelocity} based on $\mathbf{u}(t + \Delta t)$ and  $\mu(t + \Delta t)$;
\item Update the position of the boundary nodes as detailed in Section \ref{growth} to obtain $\mathcal{R}^{\rm R}(t + \Delta t)$;
\item Remesh the new domain $\mathcal{R}^{\rm R}(t + \Delta t)$ to obtain the new computational domain;
\item Reinterpolate the newly computed swelling ratio $J(t + \Delta t)$ on the updated computational domain $\mathcal{R}^{\rm R}(t + \Delta t)$;
\item Proceed to next time step.
\end{enumerate}
\caption{Update of displacement $\mathbf{u}$, chemical potential $\mathbf{\mu}$, and computational domain $\mathcal{R}$ at every time step.} 
\label{alg:decoupling}
\end{algorithm}

\subsection{Initialization of the solution procedure} \label{subsec:initialization}

The iterative procedure described in Section \ref{subsec:time-semi-discretization} 
and summarized in Algorithm \ref{alg:decoupling} has to be initialized with an initial condition. Finding the initial
equilibrium configuration is a non trivial task, which requires a preliminary iterative scheme that we discuss in this section. 

As a first step, the boundary and initial conditions must be prescribed. A characteristic of the growth surface is the density of solid material formed on it, which can be determined, for example, by the distribution of chemical binding sites. Hence, the in-plane stretch components $(\bar\lambda_1, \bar\lambda_2)=(\bar\lambda_1, \bar\lambda_2)(X_1,X_2)$  must be given as a boundary condition. For the present simulation we consider a uniform and isotropic in-plane stretch $\bar\lambda_1=\bar\lambda_2=\lambda_0$ and thus the in-plane components of the boundary constraint in \eqref{u0_bc} are given by $\bar{u}_1={\left(\lambda_0-1 \right)} X_1$ and $\bar{u}_2={\left(\lambda_0-1 \right)} X_2$, while the out of plane  displacement is determined to accommodate the growth, such that $\bar u_3=-X_3$, and thus $x_3=0$ on the growth surface. Here the symmetry surface is defined along $X_3$ and $x_3$.  

\begin{figure}[H]
  \centering
    \includegraphics[width = 0.9\textwidth]{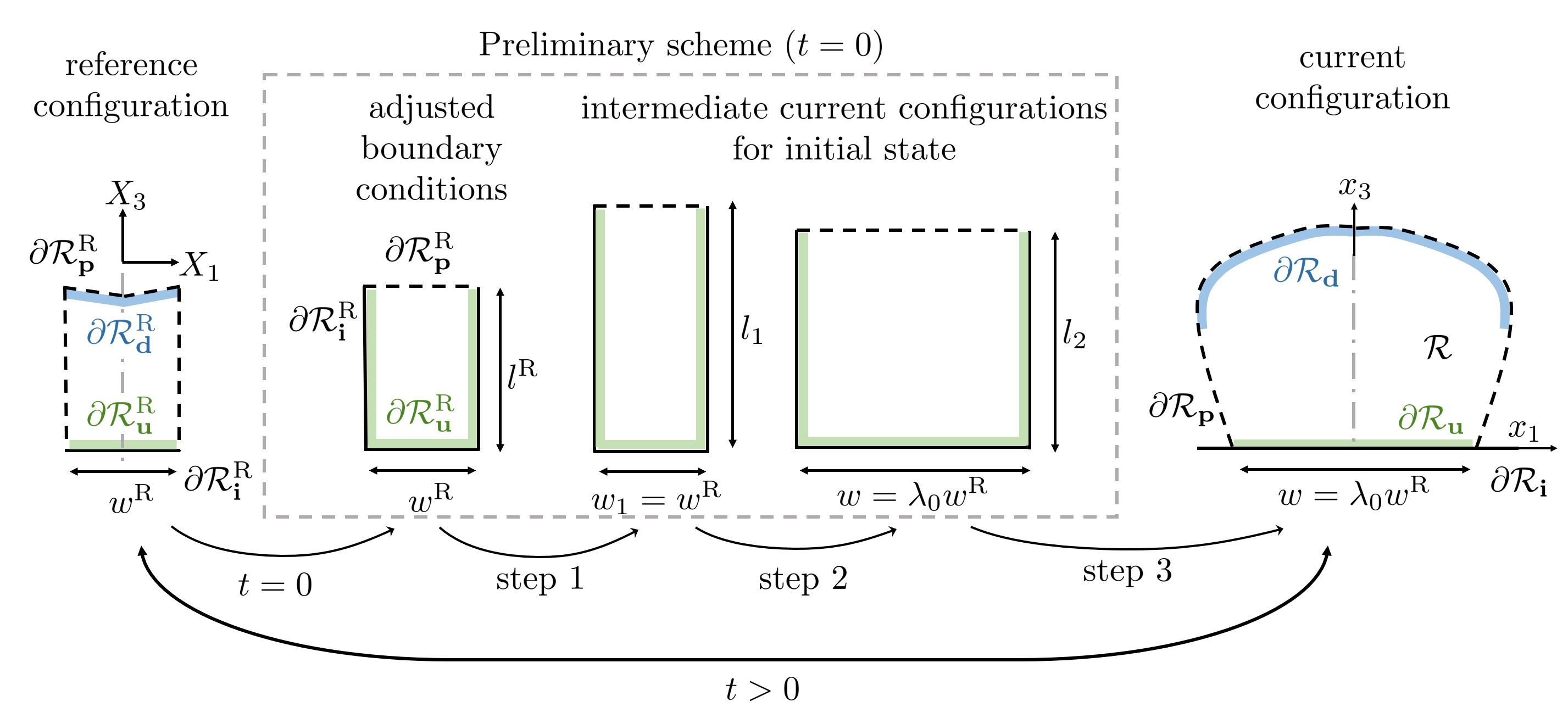}
  \caption{Illustration of the problem setting in current and reference frames, and preliminary scheme for derivation of initial current configuration via three steps. In the initial reference configuration the layer has a dry length of $l^{\rm R}$ and the dry width of the association surface is $w^{\rm R}$. The current width is $w=\lambda_0 w^{\rm R}$. Through the growth process the width of the association surface is held constant in both frames, while $l^{\rm R}=l^{\rm R}(X_1,t)$.}
  \label{fig2}
\end{figure}

Due to initial swelling and the constraint to the substrate, the initial configuration itself is not a trivial homogeneous state, and thus  a  preliminary  numerical scheme is applied to recover the initial state before allowing growth to commence. To achieve convergence of the strongly coupled and highly nonlinear system, the preliminary scheme is divided into three steps, as illustrated in Fig. \ref{fig2}. The \textit{first step} initiates from the homogeneous dry state with $\textbf{F} = \textbf{I}$ and with the simplified boundary conditions adapted to represent the uniaxial field  by replacing the chemical potential boundary condition \eqref{mu_cont} with a  displacement boundary condition on $\partial \mathcal{\rm R}_{\textbf p} \setminus \partial \mathcal{\rm R}_{\textbf d}$ to allow motion   only along $x_3$. We then proceed in a step-wise manner to accommodate swelling, where each integration step takes as initial guess the result from the previous step\footnote{Note that in the preliminary scheme, depending on the number of integration steps, the diffusion need  not be fully equilibrated.}. Further, this initial step solves only a linearized version of the constitutive equations (as described in the Appendix). In the \textit{second step}, the same system is solved with the fully nonlinear equations while $\lambda_1$ and $\lambda_2$ are  raised incrementally to the target value $\lambda_{0}$. Finally, in the \textit{third step} the appropriate boundary conditions are used to obtain the solution for the full 2D geometry. 

Once the preliminary scheme is complete, we proceed to perform the time integration for the growth process, as described in Section \ref{subsec:time-semi-discretization}. Note that in contrast to  kinematic growth theories, here no intermediate growth configuration is needed, as both the current and reference configurations evolve simultaneously\footnote{An intermediate configuration is often used in kinematic growth theories via multiplicative decomposition of the deformation gradient \citep{Menzel}.}. 

\section{Results}\label{sec:sim}

In this section, we apply the presented numerical growth simulation framework to four different settings. First, we verify the numerical framework and compare our results with the analytical results obtained by \cite{abiakl} for uniaxial growth. Then, we proceed to describe growth in two dimensions, varying the substrate properties, as defined by the stretch $\lambda_0$ and the width of the association surface $w^{\rm R}$. 
All other parameter values are kept constant in the presented simulations. 

The model is built on the two open-source software components: Fenics, a Finite Element Method (FEM) software library \citep{Fenics-General,AlnaesBlechta2015a}; and Gmsh, a 3D finite-element mesh generator \citep{Gmsh}. The output is then processed using the open-source large data visualization tool Paraview \citep{Paraview}.

\subsection{Uniaxial growth}\label{subsec:1D}
The theoretical model detailed in Section \ref{sec:model} can be solved analytically in simplified geometric settings. \cite{abiakl} obtained an analytical solution  for growth on a flat substrate where the association surface is unbounded in the $X_1$ and $X_2$ directions such that growth occurs only along $X_3$. Later,   this analytical solution was extended to account for the effect of curvature and  substrate  deformation, by considering radial growth on deformable spherical substrates \citep{abi2020surface}. In all cases, it was found that following a rapid diffusion dominated response, the growth process tends towards a \textit{universal path} that is independent of initial conditions. Along this universal path growth and diffusion act harmoniously as the system evolves towards a \textit{treadmilling} state. In this treadmilling state, association and dissociation are balanced such that the body no longer changes its dimensions in the current frame, although association and dissociation reactions persist. In this section, as a validation of our numerical model, we consider uniaxial growth on a flat substrate and compare our results with the analytical results from \cite{abiakl}.

\begin{figure}[H]
  \center
    \includegraphics[width = 0.8\textwidth]{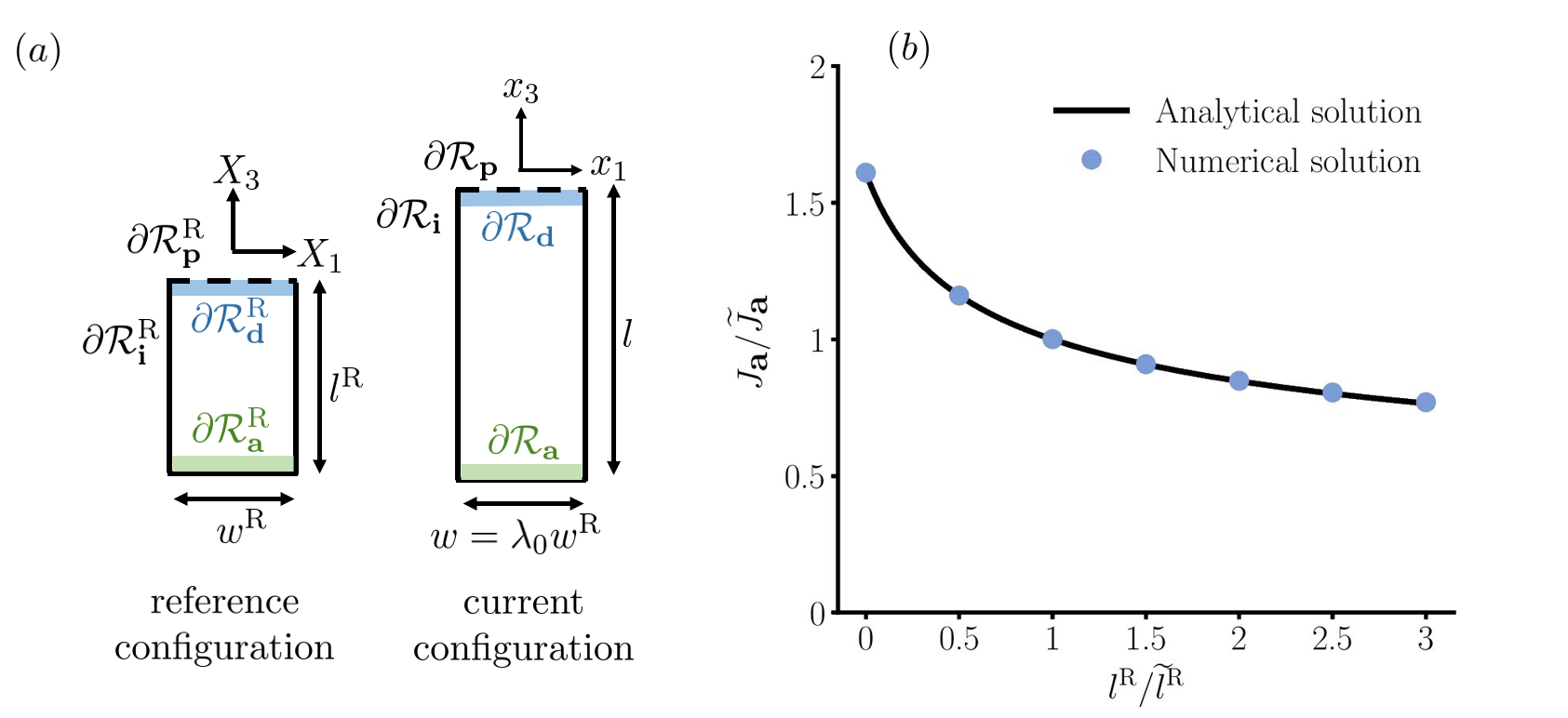}
  \caption{(a) Illustration of reference and current configurations for the case of uniaxial growth, as well as corresponding boundary conditions. Here $\partial \mathcal{R}_{\textbf{p}} = \partial \mathcal{R}_{\textbf{s}}$ and $\partial \mathcal{R}_{\textbf{i}} = \partial \mathcal{R}_{\textbf{u}}$. (b) Comparison of analytical universal path from \cite{abiakl} with numerical simulation results, showing the swelling ratio at the association surface $(J_{\textbf a})$  as a function of the dry length $(l^{\rm R})$. Both are normalized by their treadmilling values, as denoted by the superimposed tilde. These results are obtained using $\lambda_0= 4.96$. }
\label{up}
\end{figure}

To constrain the growth to occur only along the $X_3$ direction, we subject the faces of the rectangular body  with $\textbf{n}^{\rm R}=(0, 0, \pm 1)$, to the displacement boundary condition \eqref{u0_bc} such that $\bar{u}_1=\bar{u}_2=0$, and render them impermeable by applying the boundary condition \eqref{massneumann} with $\bar{j}^{\rm R}=0$, as illustrated in Fig. \ref{up}(a). We permit association  only on the substrate where $\textbf{n}^{\rm R}=(0,0,-1)$.  

Simulations for layers with different initial lengths $(l^{\rm R})$  were conducted, spanning the time it takes for diffusion to equilibrate (the width $w^{\rm R}$ does not effect these results), and are represented by the points in Fig. \ref{up}(b). It is expected that the points coincide with the universal path, which is obtained via the analytical solution, and represented by the black curve. The remarkable agreement, observed for all cases, provides a validation of our numerical scheme. In the next section we will expand on these simulations by removing the vertical constraint and allowing for two-dimensional morphologies to emerge. 

\subsection{Two-dimensional growth}\label{subsec:2D}
In the following, we present results for the evolution of a 2D growing system and discuss the sensitivity of the growth process to the two parameters $\lambda_0$ and $w^{\rm R}$, which are properties of the growth surface. The current and reference configurations in the 2D setting are illustrated in Fig. \ref{fig1} and later also in Fig. \ref{fig2}, where additional details on the geometry are provided. Here we consider growth scenarios for which we set $\partial \mathcal{R}_{\textbf{p}} = \partial \mathcal{R}_{\textbf{s}}$ and $\partial \mathcal{R}_{\textbf{i}} = \partial \mathcal{R}_{\textbf{u}}= \partial \mathcal{R}_{\textbf{a}}$. In contrast to the uniaxial setting, described in the previous section, flux is now permitted through the $\textbf{n}^{\rm R}=(0,\pm 1,0)$ faces, which are no longer geometrically constrained.  

\begin{figure}[H]
\centering
        \includegraphics[width=\linewidth]{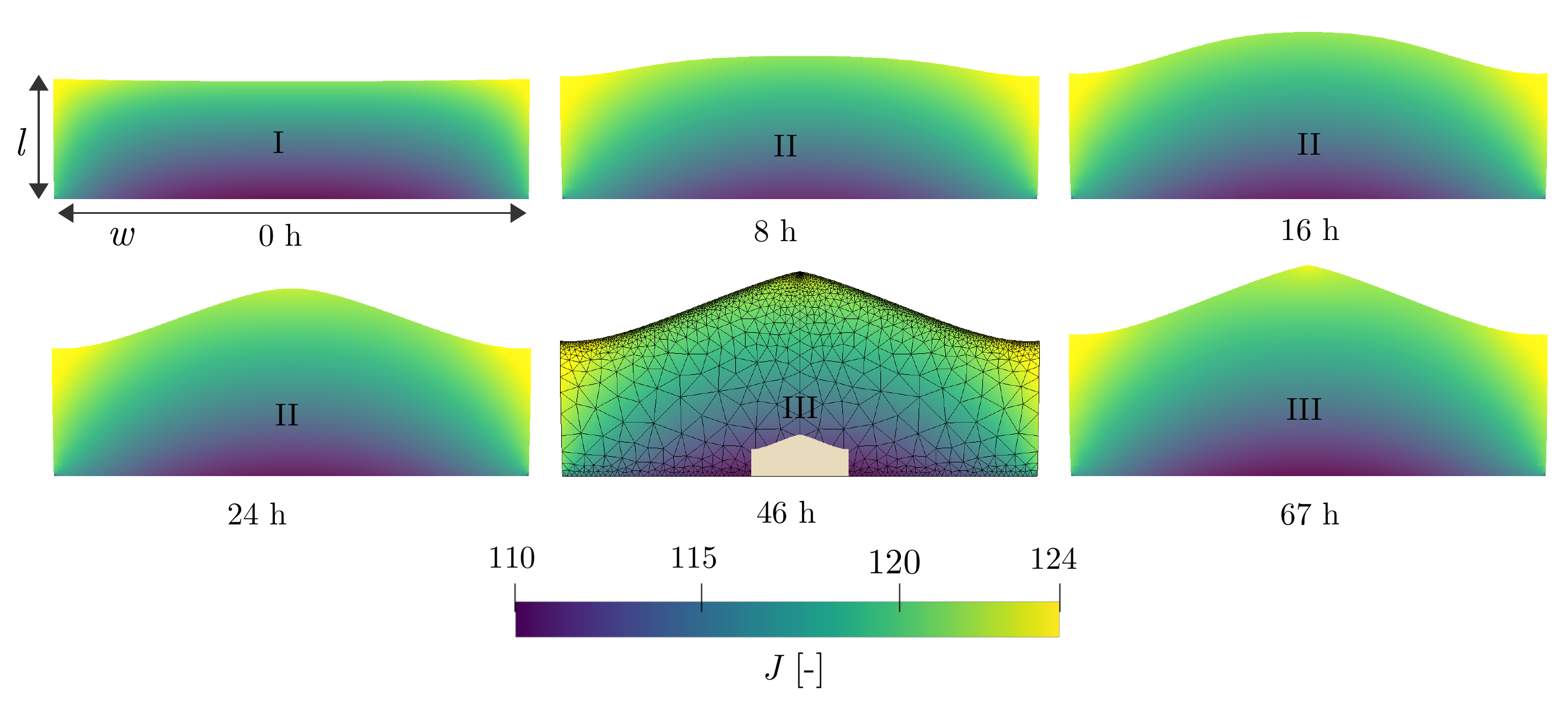}
    \caption{Results for $w^{\rm R} = \SI{2}{mm}$ and $\lambda_0 = 4.88$ in the current configuration, with $w=\lambda_0 w^{\rm R}$ and the thickness $l$, at the edge evolving from $ \SI{2.5}{mm}$ to $ \SI{2.8}{mm}$. At time t = \SI{46}{h}, the mesh and the reference configuration are included in the image. (I) denotes the diffusion-dominated stage, (II) the morphogenesis stage, and (III) the proportionate growth stage. Here the color map represents the swelling ratio $J$, corresponding images portraying changes in chemical potential are shown in Fig. \ref{img:mu} of the Appendix. A video of the simulation is provided  in the \textcolor{blue}{\href{https://www.dropbox.com/s/j2hz1lmuup0rvw8/488_2mm.mp4?dl=0}{supplementary material}.}
    }\label{247}
\end{figure}

A representative simulation of the growth process in the current frame is shown in Fig. \ref{247} for a sample with $w^{\rm R} = \SI{2}{mm} $, and  $\lambda_0 = 4.88$. At $t=0$ the swollen sample begins to associate/dissociate at its boundaries. Due to the transformation of solvent into solid material at the association surface, the solvent content and hence the swelling ratio $J$ is smallest in that region. At the permeable boundary, continuity of chemical potential leads to a high solvent content and a larger swelling ratio $J$, with the maximum value occurring at the corners. The shape that is developed over time  depends on the fully coupled response of the material system. 
As observed, the net association is maximum at the mid-section, leading to thickening there, while the net association appears to be minimal near the edges. This is due to the competition between association and dissociation, and is highly affected by both the local fields and the orientation of the surface normal, which act together to define the final morphology of the growing body. A representative mesh (recall that a new mesh is formed at every time step) is shown for $t=\SI{46}{h}$, as well as the corresponding reference configuration at that time. The large deformations that occur between the two configurations lead to  significant nonlinear effects that are fully captured by the present framework. 

In examining the growth process %\textcolor{red}{(a video of the entire simulation is shown also in supplementary material)} \textcolor{blue}{\href{https://www.dropbox.com/s/qrtl2mtenhsbbmj/4907_1mm.mp4?dl=0}{Link to video}} 
we identify four stages: (I) a \textit{diffusion-dominated} stage, in which large and rapid adaptations in swelling, $J$, occur in response to the sudden change in boundary conditions, (II) a \textit{morphogenesis} stage, in which  changes in morphology are observed while relatively small changes in swelling occur, (III) a  \textit{proportionate growth} stage, in which the morphology of the dissociation surface  and the swelling field, $J$, are nearly unchanged while the dimension of the body continues to evolve by homogeneous addition and removal of material at the association/dissociation surfaces, and (IV) a \textit{treadmilling} state, in which both the morphology and the spatial fields remain constant in time. The last stage is approached asymptotically, and will be discussed in more detail in the next paragraph.  In contrast to uniaxial growth, which, as explained in the previous section, consists of three stages, the  morphogenesis stage emerges here due to the additional degree of freedom in the deformation field, whereas the proportionate growth stage is a  2D analogue of the universal path, which has been observed in the uniaxial settings.

\begin{figure}[H]
\centering
        \includegraphics[width=0.9\linewidth]{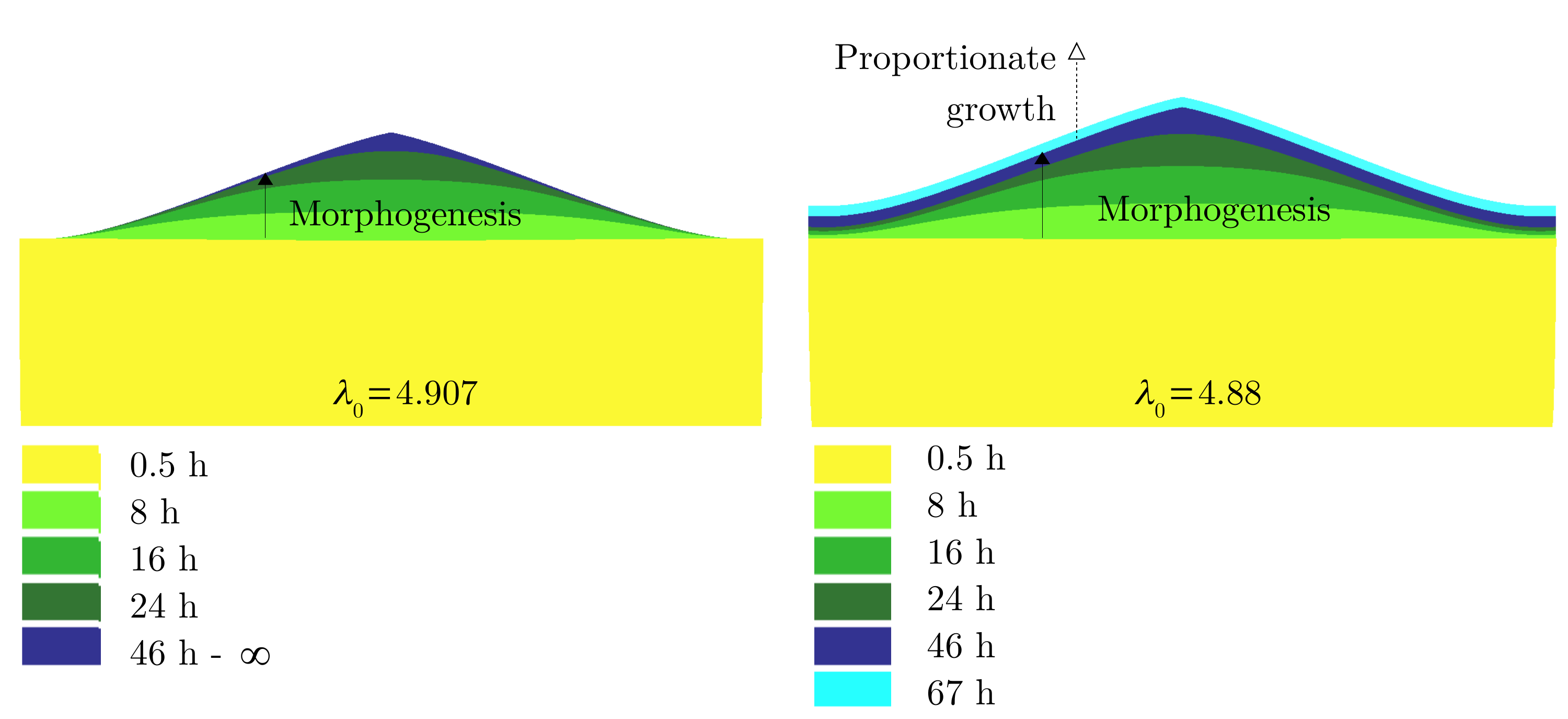}
    \caption{
    Comparison of results for $\lambda_0 = 4.907$ against $\lambda_0 = 4.88$, obtained with $w^{\rm R} = \SI{2}{mm}$ and shown in the current configuration.
    }\label{growth_stages}
\end{figure}

The different stages of  growth (i.e. stages I-III)  have been consistently observed in all of our simulations and are further  clarified in Fig. \ref{growth_stages}, by comparing the morphological evolution of the body shown in Fig. \ref{247} with that of a body grown on a substrate with the same width  $w^{\rm R} = \SI{2}{mm}$, but with a slightly different imposed stretch   $\lambda_0 = 4.907$. %(see corresponding video in the \textcolor{blue}{\href{https://www.dropbox.com/s/omajgz1sc8l0ymx/4907_2mm.mp4?dl=0}{supplementary material}} and Fig. \ref{img:J4907} in the Appendix)
In both cases, the morphology that emerges after the diffusion dominated stage is shown to be nearly rectangular (yellow regions). Morphogenesis then dominates and occurs for approximately $t\sim \SI{46}{h}$ in both samples. It results in significant thickening of the layer in the mid region, with minimal net growth near the edges. Interestingly, following this morphogenesis, the layer with $\lambda_0 = 4.907$ appears to directly  approach the treadmilling state, while the layer with $\lambda_0 = 4.88$ continues to thicken for the duration of the simulation, but without any appreciable changes in its morphology. The time required for this proportionate growth stage to approach treadmilling is determined by the chosen initial thickness $l^{\rm R}$. For the sample with   $\lambda_0 = 4.907$, this initial thickness  matches the final treadmilling thickness at the edges, thus significantly reducing the time needed to arrive at full treadmilling, by factually skipping the proportionate growth stage. 
The same initial thickness for the sample with $\lambda_0 = 4.88$ first transcends  into the proportionate growth stage at $46$ h and remains in that stage for the rest of the simulated timeframe, with approximately constant net growth velocity $\dot{l}^{\rm R}$.  It can be assumed that  this sample will also reach treadmilling eventually. However the evolution up to this point occurs on a timescale much larger than the one investigated in this work. The amount of time it takes to complete morphogenesis is the same for both samples, however the path of the two settings differs largely once the shape is formed. This result also exemplifies the significant sensitivity of the growth process to  the imposed stretch, $\lambda_0$. 

\newpage
Quite remarkably, the four stages of growth observed in our simulations are reminiscent of  growth  processes in the natural world, and yet emerge here spontaneously without any externally imposed regulation or coordination. Skin repair, for example, consists of a sequence of well defined phases \citep{lucas2010differential,sorg2017skin}:  the early stage is characterized by an immediate influx of growth factors; this is followed by the formation of tissue that fills the wound; then, when the wound space is refilled and the epidermis is restored, the final stage, consists of tissue maturation. These three phases are analogous to stages I-III in our simulations, while the final healthy result is the treadmilling state (IV). 

\begin{figure}[H]
    \centering 
    \includegraphics[width=\linewidth]{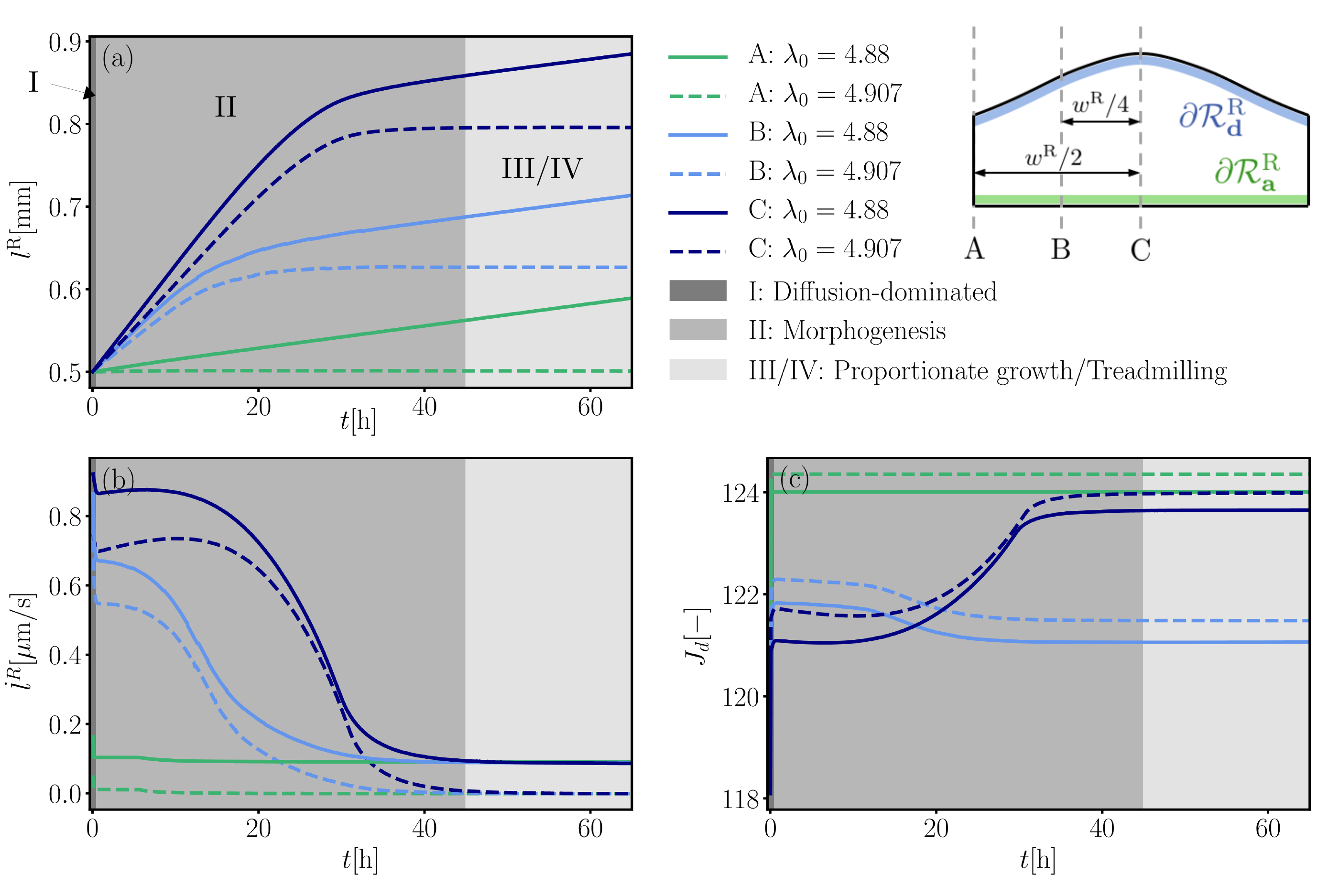} 

  \caption{Sensitivity to prestretch $\lambda_0$ is shown by examining the evolution of (a) the dry length  $l^{\rm R}$, (b) its rate of change $\dot{l}^{\rm R}$, and (c) the swelling ratio at the dissociation surface $J_{\rm d}$, for the simulations shown in Figs. \ref{247} and \ref{growth_stages} at three different locations, denoted by A, B, C as illustrated in the inset. }
  \label{img:lambda_sensitivity}  
\end{figure}

To better understand the sensitivities of the growth process, the results in Figs. \ref{247} and \ref{growth_stages}  are further examined in Fig. \ref{img:lambda_sensitivity} showing the evolution of the normalized dry length $l^{\rm R}$, its rate $\dot{l}^{\rm R}$, and the swelling ratio at the dissociation surface $J_\text{\rm d}$, for two different values of the prestretch $\lambda_0$, and at three different locations; A - the edge, B - the mid line, and C - the symmetry surface (as illustrated). Both configurations undergo a fast diffusion-dominated stage I in the first \SI{0.5}{h} before they transcend into the morphogenesis stage II. After $\SI{46}{h}$, the simulation with $\lambda_0 = 4.88$ arrives at the proportionate growth stage III, where for the duration of the simulation the value of $\dot{l}^{\rm R}$ remains approximately constant. It can be assumed that, if significantly longer simulation times are considered, it will further develop towards treadmilling. The material with larger prestretch $\lambda_0 = 4.907$ also completes morphogenesis at $\SI{46}{h}$, but after this directly approaches the treadmilling state (i.e. stage IV). Fig. \ref{img:lambda_sensitivity}(c) shows that after initial diffusion, $J_{\rm d}$ continues to evolve only during stage II. Since changes in the orientation of the dissociation surface directly influence the value of $J_{\rm d}$, this further confirms that morphogenesis is dominant only in stage II. Later, $J_{\rm d}$ approaches a constant value. Notice that during morphogenesis, $J_{\rm d}$  remains constant at A, it increases at the symmetry surface and becomes smaller in-between (at B).

\begin{figure}[H]
    \centering 
    \includegraphics[width = \linewidth]{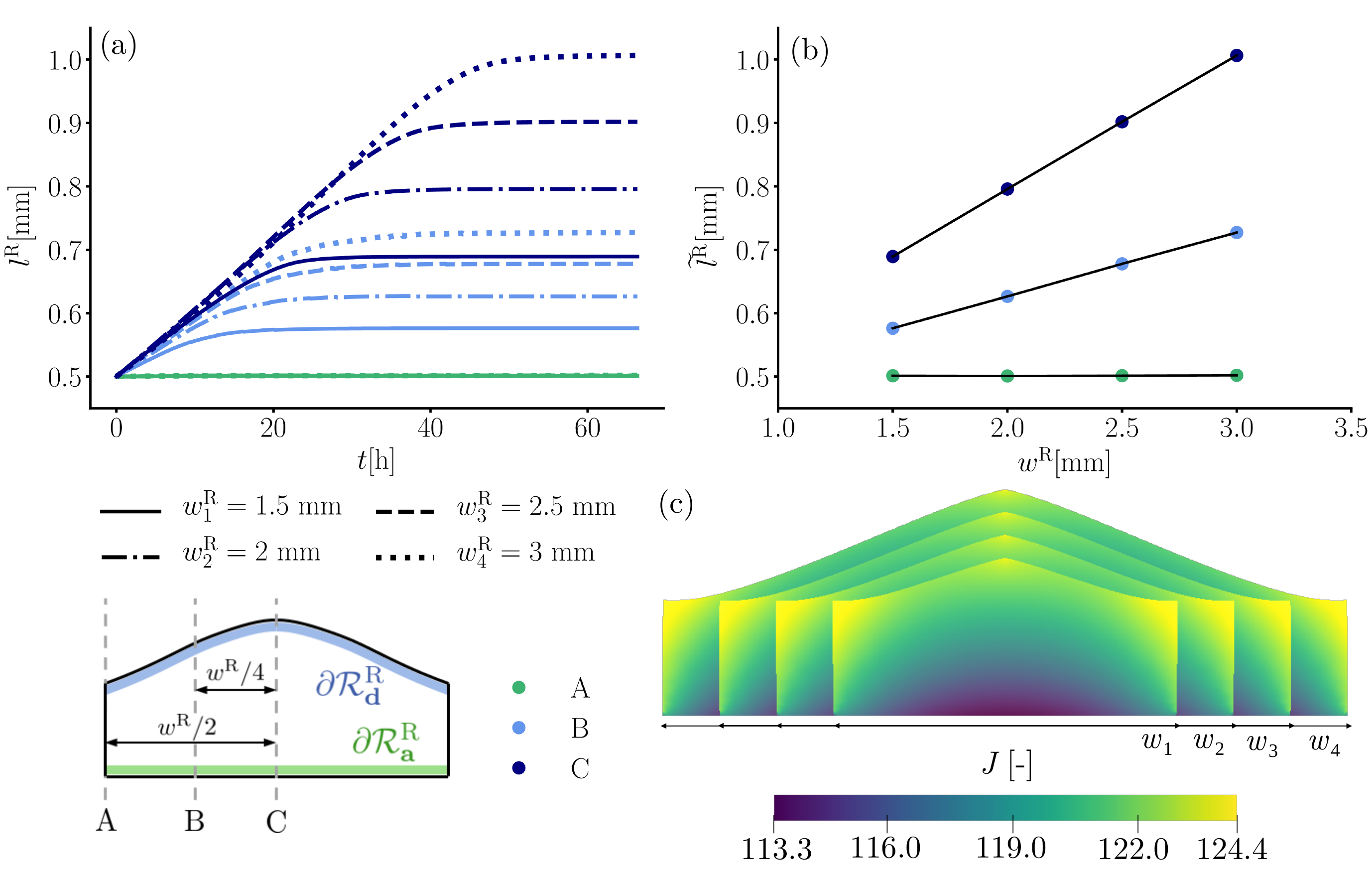}
  \caption{Material evolution for different dry widths $w^{\rm R}$ with prestretch $\lambda_0 = 4.907$ and $w = \lambda_0 w^{\rm R}$. (a) Evolution of dry length at the border (A), the mid-surface (B), and the symmetry surface (C). (b) Approximate treadmilling length as a function of $w^{\rm R}$. (c) Approximate treadmilling states, represented in the current configuration.} 
  \label{img:width_sensitivity}  
\end{figure}

Next, we investigate the sensitivity of the model to the width of the association surface $w^{\rm R}$. Figs. \ref{img:width_sensitivity}(a,b) show the evolution of dry length over time  as well as the  (approximate) treadmilling length for different values of $w^{\rm R}$. Fig. \ref{img:width_sensitivity}(c) shows the (approximate) treadmilling morphologies of bodies grown on substrates with different width $w^{\rm R}$.  A larger width of the association surface results in a larger final length, which, as shown in Fig. \ref{img:width_sensitivity}(b), exhibits a linear relation. However, since the analysis is limited to four example cases, it is possible that this observed linearity is limited to the considered region dimensions.
Finally, from these results it is evident that the width of the association surface is a well-suited tool to control the final dimensions of the grown material.

\begin{figure}[H]
\centering
        \includegraphics[width=\linewidth]{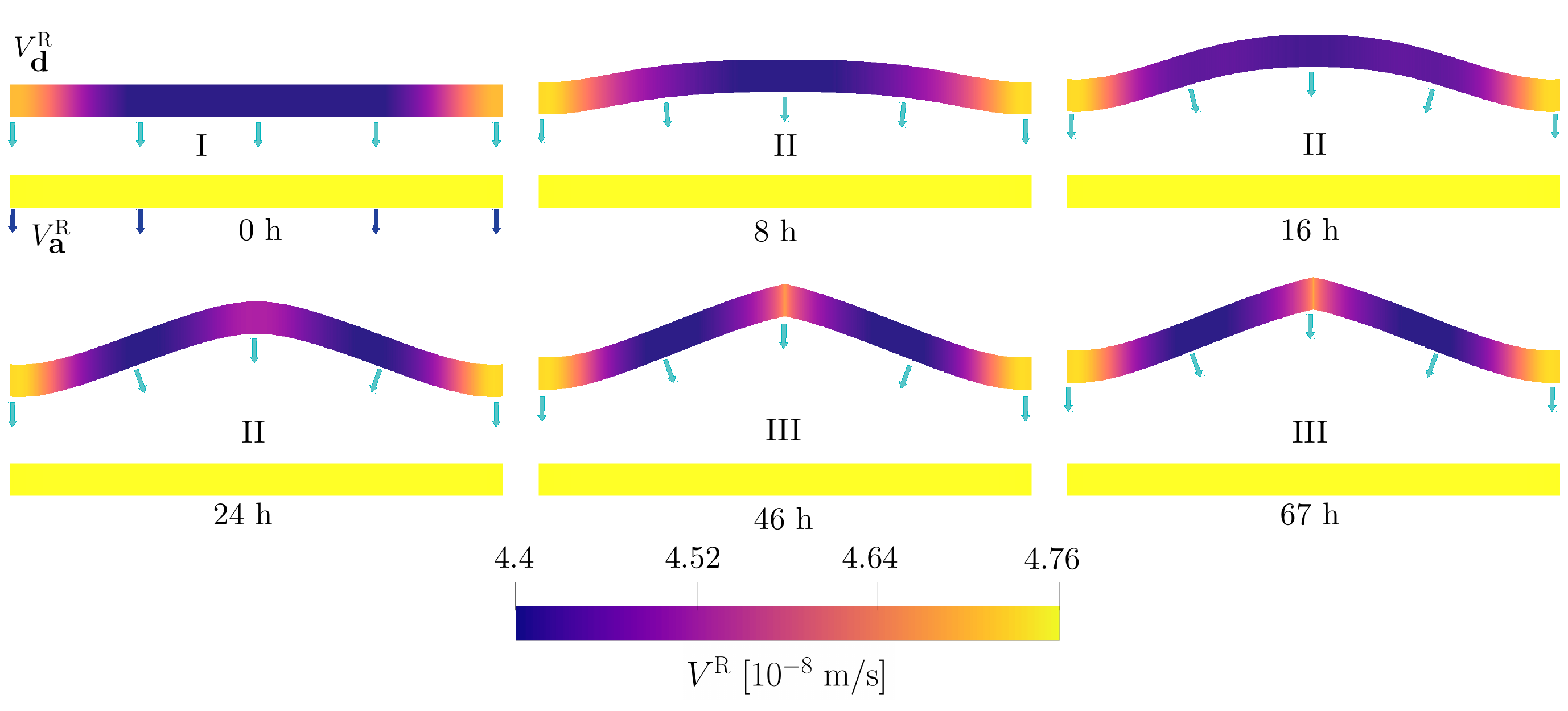}
        \caption{Evolution of the dissociation and association velocity over time in the reference configuration for $w^{\rm R}=\SI{2}{mm}$ and $\lambda_0 = 4.88$. The boundary normals of the dissociation surface $\textbf{n}^{\rm R}_\textbf{d}$ are denoted by light blue arrows at each time, while $\textbf{n}^{\rm R}_\textbf{a}$ are shown as dark blue arrows only for $\SI{0}{h}$ and remain constant over time.} \label{velocities}
\end{figure}

To further understand the dynamics that lead to the morphological and proportionate changes observed in our simulations, it is insightful to examine  the association and dissociation velocities and how they vary both with time, and along the boundaries, as illustrated in Fig. \ref{velocities}  for the system with $\lambda_0 = 4.88$ and $w^R = \SI{2}{mm}$. Notice that the material is shown in the reference configuration and the color map refers to the magnitude of the vector of the growth velocity, $V^{\rm R}$. Accordingly, in the proportionate growth stage (which starts at at $t = \SI{46}{h}$ and continues until the end of the simulation), the components perpendicular to the substrate, $V^{\rm R}_{\textbf{d},3}$ and $V^{\rm R}_{\textbf{a},3}$, are constant for both velocities along the respective boundary, yet the magnitudes of the velocity vectors are not.
%differ.

By examining the relation for the driving force of growth \eqref{driving_force}, in view of the boundary conditions on the dissociation surface (i) free boundary ($\bar{\textbf{s}}=0$) in \eqref{momentumneumann} and (ii) continuity of chemical potential \eqref{mu_cont}, it becomes clear that $V^{\rm R}_\textbf{d}\left(\textbf{u},\mu \right) \approx V^{\rm R}_\textbf{d}\left(J \right)$. In contrast, at the association surface, the spatial and temporal variations of $V^{\rm R}_\textbf{a}$ do not show similar dependence on $J$ and are negligible when compared to $V^{\rm R}_\textbf{d}$. From our numerical investigation we find that $V^{\rm R}_\textbf{a}\left(\textbf{u},\mu \right) \approx V^{\rm R}_\textbf{a}\left(\mu \right)$. The swelling ratio $J$ is shown in Fig. \ref{247} and the chemical potential $\mu$ is shown in Fig. \ref{img:mu} of the Appendix. Overall, these dependencies elucidate the  interplay between morphogenesis and the swelling on the dissociation surface, $J_d$: while morphogenesis is the main driver of the time evolution of $J_d$,  the changes in $J_d$, in turn, lead to the slowdown and eventual conclusion of morphogenesis.

\subsection{Influence of the growth regularization step}\label{regular}
The regularization step, presented in Section \ref{growth}, removes residual stresses that may be induced by differential growth along the association surface. Although this assumption becomes accurate for the limits of an infinitely wide association surface (i.e. uniaxial growth) and for a thin association surface (i.e. growth of a filament), it remains to determine its influence  on growth of finite layers, as considered in this work.

In the present system, the dry length of material that has been formed at a given point on the substrate, between the time of initiation of growth $(t=0)$ and time $t$, is $\int_0^t V^{\rm R}_{\textbf{a}}{\rm d} t$. An upper bound estimation of the differential growth   can be obtained as 
\begin{equation}
\Delta l^{\rm R}(t) \sim \max\left\{\Delta V^{\rm R}_{\textbf{a}}\right\} \cdot t,
\end{equation}
where,  $\max\left\{\Delta V^{\rm R}_{\textbf{a}}\right\} $ is taken as the maximum difference in association velocities between two points  throughout the growth process. Effects of differential growth are expected to be small if $\Delta l^{\rm R}\ll w^{\rm R}$, where $w^{\rm R}$ is used as a representative distance between the two points. 

To estimate the influence of differential growth in our system, for the discussed setting with $\lambda_0= 4.907$, we use the maximum difference in growth rate, which occurs between points $A$ and $C$ of the association surface (points located as illustrated in Figure \ref{img:width_sensitivity}), and we consider the final time of the simulation $t = 67.5$ h. 
From the simulation results we find that at point A -  $V^{\rm R}_{\textbf{a}}= \SI{4.758d-8}{m.s^{-1}}$ and at  point C - $V^{\rm R}_{\rm{a}} = \SI{4.750d-8}{m.s^{-1}}$ and thus,  $\Delta V^{\rm R}_{\textbf{a}} = \SI{8d-11}{m.s^{-1}}$. This difference is found to remain approximately constant through time. 

For all of the shown simulations, the width in the reference configuration is of the order of $ w^R \sim \SI{1}{mm}$, hence the ratio is
\begin{equation}
\Delta l^{\rm R} /    w^R \sim 0.01,
\end{equation}
Accordingly, in the present setting, the resulting residual stresses can be considered negligible. 

Although the above argument cannot capture coupled effects, which in complex growth settings can drive formation of residual stresses through a positive feedback loop, it  provides a strong indication that residual stresses have a negligible effect on  growth of finite layers with the considered material properties. Further advancement of the numerical method is required to determine the limits of this assumption and to consider growth on nonuniform substrates.

\section{Concluding remarks}\label{sec:conclusion}
In this paper, we developed a numerical framework for simulating the evolution of bodies that can grow and dissociate due to chemical reactions that occur on their boundaries. The presented framework captures the coupled  effects of  large deformations and diffusion on  the response in the bulk and on the  kinetics that drive the boundary reactions. 
Three main ingredients are incorporated to facilitate the numerical integration in both space and time: (i) The coupling between the nonlinear diffusion and deformation fields in the bulk is treated via  finite element discretization; it is initiated for given initial body dimensions via a specialized scheme and then solved using an iterative time-marching method consisting of a Newton-Raphson linearization. (ii) The rates of  association and dissociation determined by the kinetic relation are used 
to determine the evolution of the integration domain and simplified via a volume preserving regularization step, which is shown to provide accurate results in the considered  material system. (iii) The new domain is updated explicitly  and remeshed upon convergence of the Newton-Raphson iterations at every time step. 
Upon validation of the numerical framework in comparison with analytical results obtained for a uniaxial growth setting, we present simulations of growth of finite bodies in a generalized plane-strain setting and exam sensitivities of the growth process to properties of the association surface. 
From these simulations, four distinct stages of growth emerge;  a rapid diffusion-dominated stage is followed by a morphogenesis stage, then a proportionate growth stage appears and exhibits  minimal changes in shape, until a  treadmilling state is attained. These growth stages that spontaneously appear in our simulations which include only the basic mechanisms by which material systems may grow and evolve, exhibit a striking similarity to growth processes that occur in nature, hence  these results can potentially help interpret observations of growth. While the present approach is shown to apply for growth of finite bodies on  flat uniform substrates, future work should center on extension of this model to arbitrary geometrical settings.

\section*{Appendix }
\newcounter{defcounter}
\setcounter{defcounter}{0}
\newenvironment{myequation}{%
        \addtocounter{equation}{-1}
        \refstepcounter{defcounter}
        \renewcommand\theequation{A\thedefcounter}
        \begin{equation}}
{\end{equation}}

\setcounter{figure}{0}
\renewcommand{\thefigure}{A\arabic{figure}}

\setcounter{subsection}{0}
\renewcommand{\thesubsection}{A.\arabic{subsection}}

\subsection{Linearization}

As described in Section \ref{subsec:initialization}, the configuration that the iterative procedure is initialized with is nonhomogeneous, due to initial swelling subjected to the substrate constraint. To obtain this configuration, a  preliminary numerical scheme facilitates convergence by using a linearized version of the constitutive relations, as detailed next. 
 To simplify the momentum
balance equation we write the linearized form of \eqref{eq:pressure},  using the following Taylor expansion
\begin{myequation}\label{expansion}
\ln \left( 1 - x \right) = -\sum_{n=0}^\infty \frac{1}{n+1} x^{n+1}, 
\end{myequation}
which is valid for $0\leq x<1$. Then, taking $x=1/J$ in \eqref{expansion} yields
\begin{myequation}\label{Jexp}
\ln\left( 1 - \frac{1}{J} \right) = -\frac{1}{J} - \frac{1}{J^2} - \mathcal{O} \left( \frac{1}{J}\right)^3.
\end{myequation}

Under the assumption that $J \gg 1$, higher order terms can be neglected and this finally leads to the  expression
\begin{myequation}
p = \mu - \mu_0 - \frac{kT}{\nu}\left( \frac{\chi-0.5}{J^2}\right),
\end{myequation}
which can then be inserted into equation \eqref{eq:stresslaw}.

To simplify the evaluation of mass balance \eqref{massbalance}, the reduced form of the flux becomes
\begin{myequation}
\textbf{j}^{\rm R}= -\frac{D\nu}{kT} \text{Grad} \mu.
\end{myequation}
Next, we simplify the  energy of mixing \eqref{solventenergy} using \eqref{expansion}, which after  dropping higher order terms and inserting the relation $\phi_{\rm R}=J-1$  reads

\begin{myequation}
\psi_s(J)=(J-1) \, \psi_0+\frac{kT}{\nu} \left(\chi-1-\frac{\chi}{J} + \frac{1}{J^2} \right) 
\end{myequation}
which is then inserted into equation \eqref{growthvelocity} as part of the driving force $f$.

\subsection{Chemical potential field}
\begin{figure}[H]
    \centering 
    \includegraphics[width = \linewidth]{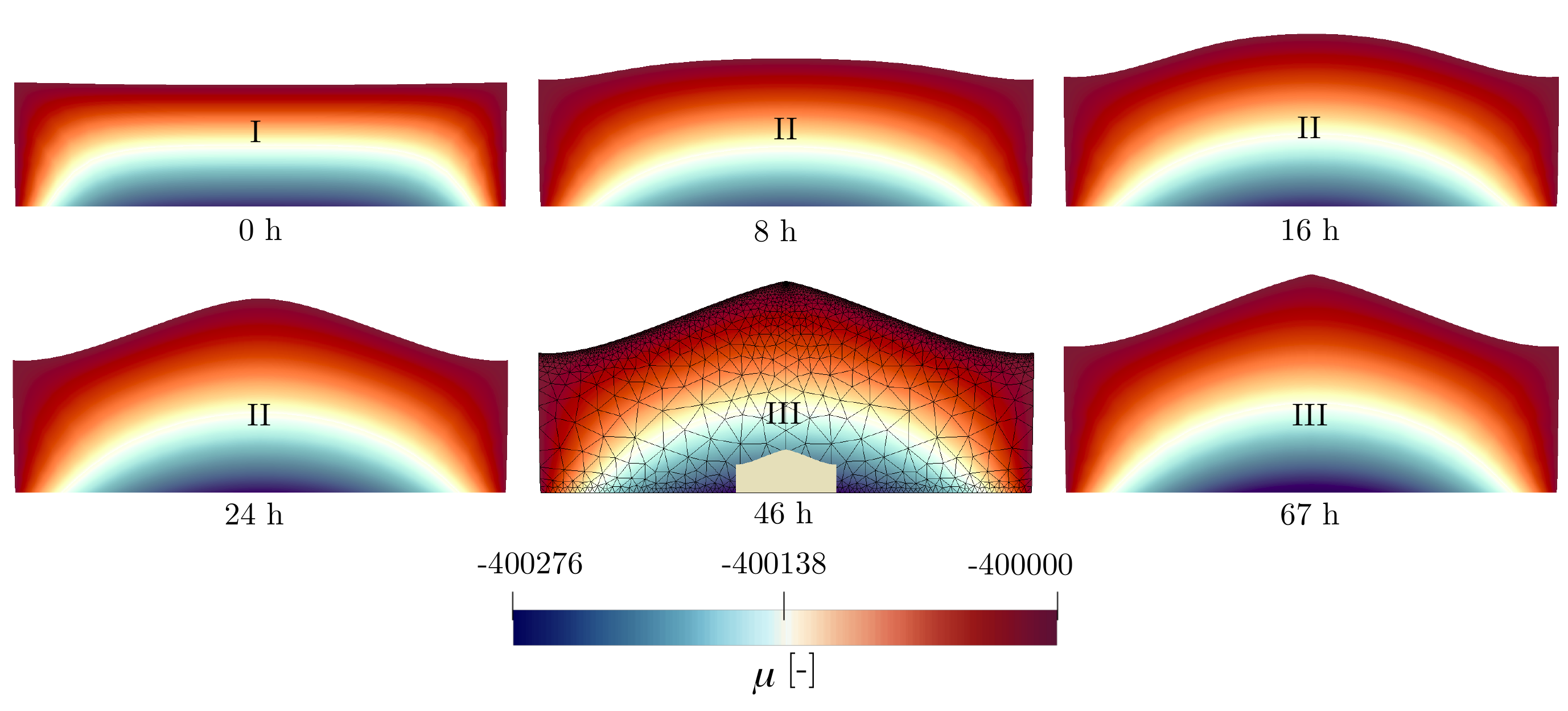}
  \caption{Evolution of the chemical potential for $w^{\rm R}=\SI{2}{mm}$ with prestretch $\lambda_0 = 4.88$ in the current configuration. (I) Diffusion-dominated stage, (II) morphogenesis, (III) proportionate growth.} 
  \label{img:mu}  
\end{figure}

Fig. \ref{img:mu} illustrates the evolution of the chemical potential from diffusion to treadmilling. Note that, due to the chemical potential boundary condition \eqref{mu_cont}, $\mu$ is constant along the dissociation boundary. The transformation of solvent to solid material at the association boundary leads to a reduced chemical potential, with the minimum value occurring at the midpoint. Unlike volume change $J$, the chemical potential $\mu$ does not undergo a significant evolution on the association or dissociation surface once diffusion has equilibriated after 0.5 h.

\subsection{Evolution up to treadmilling}
For the sample with $\lambda_0=4.907$ and initial thickness $l^{\rm R}=0.5$ mm, treadmilling is achieved immediately after morphogenesis, thus circumventing the proportionate growth stage. 

\begin{figure}[H]
    \centering 
    \includegraphics[width = \linewidth]{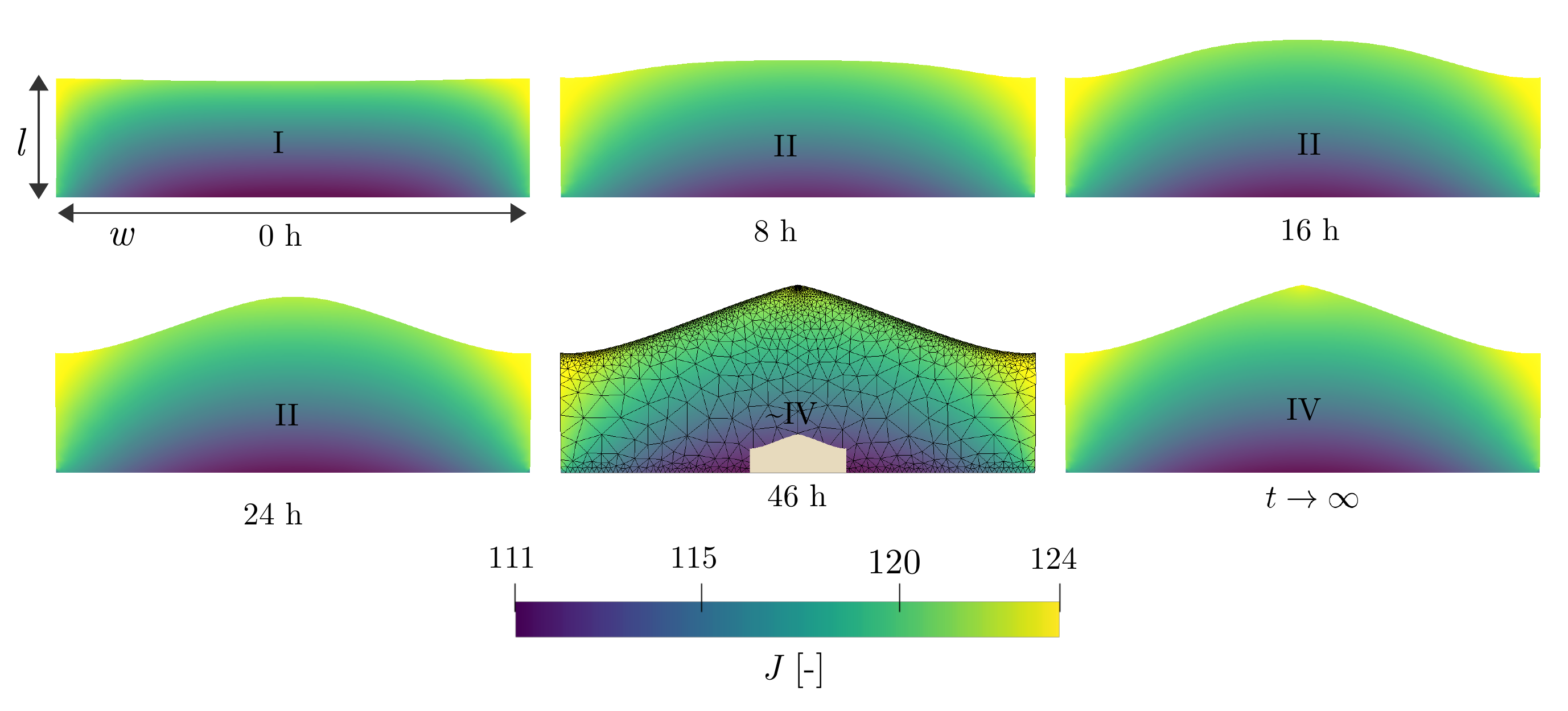}
  \caption{Evolution of the swelling ratio, $J$ for $w^{\rm R}=\SI{2}{mm}$ with prestretch $\lambda_0 = 4.907$ in the current configuration. Different stages of growth are shown: (I) Diffusion-dominated stage, (II) morphogenesis, and (IV) treadmilling, approached at $t> 46$ h. The corresponding video is provided in the \vs{\href{https://www.dropbox.com/s/omajgz1sc8l0ymx/4907_2mm.mp4?dl=0}{supplementary material}}.
  } 
  \label{img:J4907}  
\end{figure}

%\section*{\refname}
%\bibliographystyle{elsarticle-harv}
%\biboptions{sort&compress} %\bibliographystyle{unsrt}
%\bibliographystyle{numcompress} 
%\bibliography{mybibfile}

\begin{thebibliography}{59}
	\expandafter\ifx\csname natexlab\endcsname\relax\def\natexlab#1{#1}\fi
	\providecommand{\url}[1]{\texttt{#1}}
	\providecommand{\href}[2]{#2}
	\providecommand{\path}[1]{#1}
	\providecommand{\DOIprefix}{doi:}
	\providecommand{\ArXivprefix}{arXiv:}
	\providecommand{\URLprefix}{URL: }
	\providecommand{\Pubmedprefix}{pmid:}
	\providecommand{\doi}[1]{\href{http://dx.doi.org/#1}{\path{#1}}}
	\providecommand{\Pubmed}[1]{\href{pmid:#1}{\path{#1}}}
	\providecommand{\bibinfo}[2]{#2}
	\ifx\xfnm\relax \def\xfnm[#1]{\unskip,\space#1}\fi
	%Type = Article
	\bibitem[{Abeyaratne et~al.(2020)Abeyaratne, Puntel and
		Tomassetti}]{abeyaratne2020treadmilling}
	\bibinfo{author}{Abeyaratne, R.}, \bibinfo{author}{Puntel, E.},
	\bibinfo{author}{Tomassetti, G.}, \bibinfo{year}{2020}.
	\newblock \bibinfo{title}{Treadmilling stability of a one-dimensional actin
		growth model}.
	\newblock \bibinfo{journal}{International Journal of Solids and Structures}
	\bibinfo{volume}{198}, \bibinfo{pages}{87--98}.
	\newblock \DOIprefix\doi{10.1016/j.ijsolstr.2020.04.009}.
	%Type = Article
	\bibitem[{Abi-Akl et~al.(2019a)Abi-Akl, Abeyaratne and Cohen}]{abiakl}
	\bibinfo{author}{Abi-Akl, R.}, \bibinfo{author}{Abeyaratne, R.},
	\bibinfo{author}{Cohen, T.}, \bibinfo{year}{2019}a.
	\newblock \bibinfo{title}{Kinetics of surface growth with coupled diffusion and
		the emergence of a universal growth path}.
	\newblock \bibinfo{journal}{Proceedings of the Royal Society A}
	\bibinfo{volume}{475}, \bibinfo{pages}{20180465}.
	\newblock \DOIprefix\doi{10.1098/rspa.2018.0465}.
	%Type = Article
	\bibitem[{Abi-Akl and Cohen(2020)}]{abi2020surface}
	\bibinfo{author}{Abi-Akl, R.}, \bibinfo{author}{Cohen, T.},
	\bibinfo{year}{2020}.
	\newblock \bibinfo{title}{Surface growth on a deformable spherical substrate}.
	\newblock \bibinfo{journal}{Mechanics Research Communications}
	\bibinfo{volume}{103}, \bibinfo{pages}{103457}.
	\newblock \DOIprefix\doi{10.1016/j.mechrescom.2019.103457}.
	%Type = Article
	\bibitem[{Abi-Akl et~al.(2019b)Abi-Akl, Ledieu, Enke, Cordero and
		Cohen}]{biopolymer}
	\bibinfo{author}{Abi-Akl, R.}, \bibinfo{author}{Ledieu, E.},
	\bibinfo{author}{Enke, T.N.}, \bibinfo{author}{Cordero, O.X.},
	\bibinfo{author}{Cohen, T.}, \bibinfo{year}{2019}b.
	\newblock \bibinfo{title}{Physics-based prediction of biopolymer degradation}.
	\newblock \bibinfo{journal}{Soft Matter} \bibinfo{volume}{15},
	\bibinfo{pages}{4098--4108}.
	\newblock \DOIprefix\doi{10.1039/C9SM00262F}.
	%Type = Inbook
	\bibitem[{Ahrens et~al.(2005)Ahrens, Geveci and Law}]{Paraview}
	\bibinfo{author}{Ahrens, J.}, \bibinfo{author}{Geveci, B.},
	\bibinfo{author}{Law, C.}, \bibinfo{year}{2005}.
	\newblock \bibinfo{title}{Visualization Handbook}.
	\bibinfo{publisher}{Elsevier}. chapter \bibinfo{chapter}{ParaView: An
		End-User Tool for Large Data Visualization}.
	\newblock pp. \bibinfo{pages}{717--731}.
	%Type = Article
	\bibitem[{Aln{\ae}s et~al.(2015)Aln{\ae}s, Blechta, Hake, Johansson, Kehlet,
		Logg, Richardson, Ring, Rognes and Wells}]{AlnaesBlechta2015a}
	\bibinfo{author}{Aln{\ae}s, M.S.}, \bibinfo{author}{Blechta, J.},
	\bibinfo{author}{Hake, J.}, \bibinfo{author}{Johansson, A.},
	\bibinfo{author}{Kehlet, B.}, \bibinfo{author}{Logg, A.},
	\bibinfo{author}{Richardson, C.}, \bibinfo{author}{Ring, J.},
	\bibinfo{author}{Rognes, M.E.}, \bibinfo{author}{Wells, G.N.},
	\bibinfo{year}{2015}.
	\newblock \bibinfo{title}{The fenics project version 1.5}.
	\newblock \bibinfo{journal}{Archive of Numerical Software} \bibinfo{volume}{3},
	\bibinfo{pages}{9--23}.
	\newblock \DOIprefix\doi{10.11588/ans.2015.100.20553}.
	%Type = Article
	\bibitem[{Ateshian(2007)}]{ateshian2007theory}
	\bibinfo{author}{Ateshian, G.A.}, \bibinfo{year}{2007}.
	\newblock \bibinfo{title}{On the theory of reactive mixtures for modeling
		biological growth}.
	\newblock \bibinfo{journal}{Biomechanics and modeling in mechanobiology}
	\bibinfo{volume}{6}, \bibinfo{pages}{423--445}.
	\newblock \DOIprefix\doi{10.1007/s10237-006-0070-x}.
	%Type = Article
	\bibitem[{Bischofberger and Nagel(2016)}]{bischofberger2016fluid}
	\bibinfo{author}{Bischofberger, I.}, \bibinfo{author}{Nagel, S.R.},
	\bibinfo{year}{2016}.
	\newblock \bibinfo{title}{Fluid instabilities that mimic animal growth}.
	\newblock \bibinfo{journal}{Physics today} \bibinfo{volume}{69},
	\bibinfo{pages}{70--71}.
	\newblock \DOIprefix\doi{10.1063/PT.3.3307}.
	%Type = Article
	\bibitem[{Budday et~al.(2014)Budday, Steinmann and Kuhl}]{budday2014}
	\bibinfo{author}{Budday, S.}, \bibinfo{author}{Steinmann, P.},
	\bibinfo{author}{Kuhl, E.}, \bibinfo{year}{2014}.
	\newblock \bibinfo{title}{The role of mechanics during brain development}.
	\newblock \bibinfo{journal}{Journal of the Mechanics and Physics of Solids}
	\bibinfo{volume}{72}, \bibinfo{pages}{75--92}.
	\newblock \DOIprefix\doi{10.1016/j.jmps.2014.07.010}.
	%Type = Article
	\bibitem[{Chester and Anand(2010)}]{Chester}
	\bibinfo{author}{Chester, S.A.}, \bibinfo{author}{Anand, L.},
	\bibinfo{year}{2010}.
	\newblock \bibinfo{title}{A coupled theory of fluid permeation and large
		deformations for elastomeric materials}.
	\newblock \bibinfo{journal}{Journal of the Mechanics and Physics of Solids}
	\bibinfo{volume}{58}, \bibinfo{pages}{1879--1906}.
	\newblock \DOIprefix\doi{10.1016/j.jmps.2010.07.020}.
	%Type = Article
	\bibitem[{Chester et~al.(2015)Chester, Di~Leo and Anand}]{Chester2015}
	\bibinfo{author}{Chester, S.A.}, \bibinfo{author}{Di~Leo, C.V.},
	\bibinfo{author}{Anand, L.}, \bibinfo{year}{2015}.
	\newblock \bibinfo{title}{A finite element implementation of a coupled
		diffusion-deformation theory for elastomeric gels}.
	\newblock \bibinfo{journal}{International Journal of Solids and Structures}
	\bibinfo{volume}{52}, \bibinfo{pages}{1--18}.
	\newblock \DOIprefix\doi{10.1016/j.ijsolstr.2014.08.015}.
	%Type = Article
	\bibitem[{Ciarletta et~al.(2013)Ciarletta, Preziosi and Maugin}]{Ciarletta}
	\bibinfo{author}{Ciarletta, P.}, \bibinfo{author}{Preziosi, L.},
	\bibinfo{author}{Maugin, G.}, \bibinfo{year}{2013}.
	\newblock \bibinfo{title}{Mechanobiology of interfacial growth}.
	\newblock \bibinfo{journal}{Journal of the Mechanics and Physics of Solids}
	\bibinfo{volume}{61}, \bibinfo{pages}{852--872}.
	\newblock \DOIprefix\doi{10.1016/j.jmps.2012.10.011}.
	%Type = Article
	\bibitem[{Cordero and Stocker(2017)}]{cordero2017}
	\bibinfo{author}{Cordero, O.X.}, \bibinfo{author}{Stocker, R.},
	\bibinfo{year}{2017}.
	\newblock \bibinfo{title}{A particularly useful system to study the ecology of
		microbes}.
	\newblock \bibinfo{journal}{Environmental microbiology reports}
	\bibinfo{volume}{9}, \bibinfo{pages}{16--17}.
	%Type = Article
	\bibitem[{Cross and Hohenberg(1993)}]{cross1993pattern}
	\bibinfo{author}{Cross, M.C.}, \bibinfo{author}{Hohenberg, P.C.},
	\bibinfo{year}{1993}.
	\newblock \bibinfo{title}{Pattern formation outside of equilibrium}.
	\newblock \bibinfo{journal}{Reviews of modern physics} \bibinfo{volume}{65},
	\bibinfo{pages}{851}.
	%Type = Incollection
	\bibitem[{DiCarlo(2005)}]{DiCarlo}
	\bibinfo{author}{DiCarlo, A.}, \bibinfo{year}{2005}.
	\newblock \bibinfo{title}{Surface and bulk growth unified}, in:
	\bibinfo{booktitle}{Mechanics of material forces}.
	\bibinfo{publisher}{Springer}, \bibinfo{address}{Boston, MA}, pp.
	\bibinfo{pages}{53--64}.
	\newblock \DOIprefix\doi{10.1007/0-387-26261-X_6}.
	%Type = Article
	\bibitem[{Duda et~al.(2010)Duda, Souza and Fried}]{Duda}
	\bibinfo{author}{Duda, F.P.}, \bibinfo{author}{Souza, A.C.},
	\bibinfo{author}{Fried, E.}, \bibinfo{year}{2010}.
	\newblock \bibinfo{title}{A theory for species migration in a finitely strained
		solid with application to polymer network swelling}.
	\newblock \bibinfo{journal}{Journal of the Mechanics and Physics of Solids}
	\bibinfo{volume}{58}, \bibinfo{pages}{515--529}.
	\newblock \DOIprefix\doi{10.1016/j.jmps.2010.01.009}.
	%Type = Book
	\bibitem[{Feynman et~al.(1963)Feynman, Leighton and Sands}]{Feynman}
	\bibinfo{author}{Feynman, R.P.}, \bibinfo{author}{Leighton, R.B.},
	\bibinfo{author}{Sands, M.}, \bibinfo{year}{1963}.
	\newblock \bibinfo{title}{The Feynman Lectures on Physics, vol. I}.
	\newblock \bibinfo{publisher}{Addison-Wesley}, \bibinfo{address}{Reading, MA}.
	%Type = Article
	\bibitem[{Flory(1942)}]{Flory42}
	\bibinfo{author}{Flory, P.J.}, \bibinfo{year}{1942}.
	\newblock \bibinfo{title}{Thermodynamics of high polymer solutions}.
	\newblock \bibinfo{journal}{The Journal of chemical physics}
	\bibinfo{volume}{10}, \bibinfo{pages}{51--61}.
	\newblock \DOIprefix\doi{10.1063/1.1723621}.
	%Type = Article
	\bibitem[{Flory and Rehner(1943)}]{Flory43}
	\bibinfo{author}{Flory, P.J.}, \bibinfo{author}{Rehner, J.J.},
	\bibinfo{year}{1943}.
	\newblock \bibinfo{title}{Statistical mechanics of cross-linked polymer
		networks ii. swelling}.
	\newblock \bibinfo{journal}{The Journal of Chemical Physics}
	\bibinfo{volume}{11}, \bibinfo{pages}{521--526}.
	\newblock \DOIprefix\doi{10.1063/1.1723792}.
	%Type = Article
	\bibitem[{Ganghoffer and Goda(2018)}]{ganghoffer2018}
	\bibinfo{author}{Ganghoffer, J.F.}, \bibinfo{author}{Goda, I.},
	\bibinfo{year}{2018}.
	\newblock \bibinfo{title}{A combined accretion and surface growth model in the
		framework of irreversible thermodynamics}.
	\newblock \bibinfo{journal}{International Journal of Engineering Science}
	\bibinfo{volume}{127}, \bibinfo{pages}{53--79}.
	\newblock \DOIprefix\doi{10.1016/j.ijengsci.2018.02.006}.
	%Type = Article
	\bibitem[{{Geuzaine} and {Remacle}(2009)}]{Gmsh}
	\bibinfo{author}{{Geuzaine}, C.}, \bibinfo{author}{{Remacle}, J.F.},
	\bibinfo{year}{2009}.
	\newblock \bibinfo{title}{{Gmsh: A 3-D finite element mesh generator with
			built-in pre- and post-processing facilities}}.
	\newblock \bibinfo{journal}{International Journal for Numerical Methods in
		Engineering} \bibinfo{volume}{79}, \bibinfo{pages}{1309--1331}.
	\newblock \DOIprefix\doi{10.1002/nme.2579}.
	%Type = Article
	\bibitem[{Goli et~al.(2019)Goli, Robertson, Agarwal, Pruitt, Grolman, Geubelle
		and Moore}]{goli2019frontal}
	\bibinfo{author}{Goli, E.}, \bibinfo{author}{Robertson, I.D.},
	\bibinfo{author}{Agarwal, H.}, \bibinfo{author}{Pruitt, E.L.},
	\bibinfo{author}{Grolman, J.M.}, \bibinfo{author}{Geubelle, P.H.},
	\bibinfo{author}{Moore, J.S.}, \bibinfo{year}{2019}.
	\newblock \bibinfo{title}{Frontal polymerization accelerated by continuous
		conductive elements}.
	\newblock \bibinfo{journal}{Journal of Applied Polymer Science}
	\bibinfo{volume}{136}, \bibinfo{pages}{47418}.
	%Type = Article
	\bibitem[{Holland et~al.(2013)Holland, Kosmata, Goriely and Kuhl}]{Holland}
	\bibinfo{author}{Holland, M.A.}, \bibinfo{author}{Kosmata, T.},
	\bibinfo{author}{Goriely, A.}, \bibinfo{author}{Kuhl, E.},
	\bibinfo{year}{2013}.
	\newblock \bibinfo{title}{On the mechanics of thin films and growing surfaces}.
	\newblock \bibinfo{journal}{Mathematics and Mechanics of Solids}
	\bibinfo{volume}{18}, \bibinfo{pages}{561--575}.
	\newblock \DOIprefix\doi{10.1177/1081286513485776}.
	%Type = Article
	\bibitem[{Hong et~al.(2008)Hong, Zhao, Zhou and Suo}]{Hong}
	\bibinfo{author}{Hong, W.}, \bibinfo{author}{Zhao, X.}, \bibinfo{author}{Zhou,
		J.}, \bibinfo{author}{Suo, Z.}, \bibinfo{year}{2008}.
	\newblock \bibinfo{title}{A theory of coupled diffusion and large deformation
		in polymeric gels}.
	\newblock \bibinfo{journal}{Journal of the Mechanics and Physics of Solids}
	\bibinfo{volume}{56}, \bibinfo{pages}{1779--1793}.
	%Type = Article
	\bibitem[{Horstmann et~al.(2018)Horstmann, Single and
		Latz}]{horstmann2018review}
	\bibinfo{author}{Horstmann, B.}, \bibinfo{author}{Single, F.},
	\bibinfo{author}{Latz, A.}, \bibinfo{year}{2018}.
	\newblock \bibinfo{title}{Review on multi-scale models of solid-electrolyte
		interphase formation}.
	\newblock \bibinfo{journal}{Current Opinion in Electrochemistry} .
	%Type = Article
	\bibitem[{Johnson and Tabin(1997)}]{johnson1997molecular}
	\bibinfo{author}{Johnson, R.L.}, \bibinfo{author}{Tabin, C.J.},
	\bibinfo{year}{1997}.
	\newblock \bibinfo{title}{Molecular models for vertebrate limb development}.
	\newblock \bibinfo{journal}{Cell} \bibinfo{volume}{90},
	\bibinfo{pages}{979--990}.
	%Type = Article
	\bibitem[{Lai and Hu(2018)}]{lai}
	\bibinfo{author}{Lai, Y.}, \bibinfo{author}{Hu, Y.}, \bibinfo{year}{2018}.
	\newblock \bibinfo{title}{Probing the swelling-dependent mechanical and
		transport properties of polyacrylamide hydrogels through afm-based dynamic
		nanoindentation}.
	\newblock \bibinfo{journal}{Soft matter} \bibinfo{volume}{14},
	\bibinfo{pages}{2619--2627}.
	\newblock \DOIprefix\doi{10.1039/C7SM02351K}.
	%Type = Article
	\bibitem[{Li and Mooney(2016)}]{li2016}
	\bibinfo{author}{Li, J.}, \bibinfo{author}{Mooney, D.J.}, \bibinfo{year}{2016}.
	\newblock \bibinfo{title}{Designing hydrogels for controlled drug delivery}.
	\newblock \bibinfo{journal}{Nature Reviews Materials} \bibinfo{volume}{1},
	\bibinfo{pages}{16071}.
	%Type = Article
	\bibitem[{Liu et~al.(2015a)Liu, Yang, Cao, Wang, Chen, Zhang and
		Zhang}]{liu2015dehydration}
	\bibinfo{author}{Liu, Y.}, \bibinfo{author}{Yang, X.}, \bibinfo{author}{Cao,
		Y.}, \bibinfo{author}{Wang, Z.}, \bibinfo{author}{Chen, B.},
	\bibinfo{author}{Zhang, J.}, \bibinfo{author}{Zhang, H.},
	\bibinfo{year}{2015}a.
	\newblock \bibinfo{title}{Dehydration of core/shell fruits}.
	\newblock \bibinfo{journal}{Computers \& Graphics} \bibinfo{volume}{47},
	\bibinfo{pages}{68--77}.
	%Type = Article
	\bibitem[{Liu et~al.(2015b)Liu, Zhang and Zheng}]{liu2015multiplicative}
	\bibinfo{author}{Liu, Y.}, \bibinfo{author}{Zhang, H.}, \bibinfo{author}{Zheng,
		Y.}, \bibinfo{year}{2015}b.
	\newblock \bibinfo{title}{A multiplicative finite element algorithm for the
		inhomogeneous swelling of polymeric gels}.
	\newblock \bibinfo{journal}{Computer Methods in Applied Mechanics and
		Engineering} \bibinfo{volume}{283}, \bibinfo{pages}{517--550}.
	%Type = Book
	\bibitem[{Logg et~al.(2012)Logg, Mardal, Wells et~al.}]{Fenics-General}
	\bibinfo{author}{Logg, A.}, \bibinfo{author}{Mardal, K.A.},
	\bibinfo{author}{Wells, G.N.}, et~al., \bibinfo{year}{2012}.
	\newblock \bibinfo{title}{Automated Solution of Differential Equations by the
		Finite Element Method}.
	\newblock \bibinfo{publisher}{Springer}.
	\newblock \DOIprefix\doi{10.1007/978-3-642-23099-8}.
	%Type = Article
	\bibitem[{Lucas et~al.(2010)Lucas, Waisman, Ranjan, Roes, Krieg, M{\"u}ller,
		Roers and Eming}]{lucas2010differential}
	\bibinfo{author}{Lucas, T.}, \bibinfo{author}{Waisman, A.},
	\bibinfo{author}{Ranjan, R.}, \bibinfo{author}{Roes, J.},
	\bibinfo{author}{Krieg, T.}, \bibinfo{author}{M{\"u}ller, W.},
	\bibinfo{author}{Roers, A.}, \bibinfo{author}{Eming, S.A.},
	\bibinfo{year}{2010}.
	\newblock \bibinfo{title}{Differential roles of macrophages in diverse phases
		of skin repair}.
	\newblock \bibinfo{journal}{The Journal of Immunology} \bibinfo{volume}{184},
	\bibinfo{pages}{3964--3977}.
	%Type = Article
	\bibitem[{Menzel and Kuhl(2012)}]{Menzel}
	\bibinfo{author}{Menzel, A.}, \bibinfo{author}{Kuhl, E.}, \bibinfo{year}{2012}.
	\newblock \bibinfo{title}{Frontiers in growth and remodeling}.
	\newblock \bibinfo{journal}{Mechanics research communications}
	\bibinfo{volume}{42}, \bibinfo{pages}{1--14}.
	\newblock \DOIprefix\doi{10.1016/j.mechrescom.2012.02.007}.
	%Type = Article
	\bibitem[{Metzger and Krasnow(1999)}]{metzger1999genetic}
	\bibinfo{author}{Metzger, R.J.}, \bibinfo{author}{Krasnow, M.A.},
	\bibinfo{year}{1999}.
	\newblock \bibinfo{title}{Genetic control of branching morphogenesis}.
	\newblock \bibinfo{journal}{science} \bibinfo{volume}{284},
	\bibinfo{pages}{1635--1639}.
	%Type = Article
	\bibitem[{Mogilner and Oster(1996)}]{Mogilner96}
	\bibinfo{author}{Mogilner, A.}, \bibinfo{author}{Oster, G.},
	\bibinfo{year}{1996}.
	\newblock \bibinfo{title}{Cell motility driven by actin polymerization}.
	\newblock \bibinfo{journal}{Biophysical journal} \bibinfo{volume}{71},
	\bibinfo{pages}{3030--3045}.
	\newblock \DOIprefix\doi{10.1016/S0006-3495(96)79496-1}.
	%Type = Article
	\bibitem[{Mogilner and Oster(2003)}]{Mogilner2003-Force}
	\bibinfo{author}{Mogilner, A.}, \bibinfo{author}{Oster, G.},
	\bibinfo{year}{2003}.
	\newblock \bibinfo{title}{Force generation by actin polymerization ii: the
		elastic ratchet and tethered filaments}.
	\newblock \bibinfo{journal}{Biophysical journal} \bibinfo{volume}{84},
	\bibinfo{pages}{1591--1605}.
	\newblock \DOIprefix\doi{10.1016/S0006-3495(03)74969-8}.
	%Type = Article
	\bibitem[{Moulton et~al.(2012)Moulton, Goriely and Chirat}]{Moulton}
	\bibinfo{author}{Moulton, D.E.}, \bibinfo{author}{Goriely, A.},
	\bibinfo{author}{Chirat, R.}, \bibinfo{year}{2012}.
	\newblock \bibinfo{title}{Mechanical growth and morphogenesis of seashells}.
	\newblock \bibinfo{journal}{Journal of theoretical biology}
	\bibinfo{volume}{311}, \bibinfo{pages}{69--79}.
	\newblock \DOIprefix\doi{10.1016/j.jtbi.2012.07.009}.
	%Type = Article
	\bibitem[{Noireaux et~al.(2000)Noireaux, Golsteyn, Friederich, Prost, Antony,
		Louvard and Sykes}]{Noireaux}
	\bibinfo{author}{Noireaux, V.}, \bibinfo{author}{Golsteyn, R.},
	\bibinfo{author}{Friederich, E.}, \bibinfo{author}{Prost, J.},
	\bibinfo{author}{Antony, C.}, \bibinfo{author}{Louvard, D.},
	\bibinfo{author}{Sykes, C.}, \bibinfo{year}{2000}.
	\newblock \bibinfo{title}{Growing an actin gel on spherical surfaces}.
	\newblock \bibinfo{journal}{Biophysical journal} \bibinfo{volume}{78},
	\bibinfo{pages}{1643--1654}.
	\newblock \DOIprefix\doi{10.1016/S0006-3495(00)76716-6}.
	%Type = Article
	\bibitem[{Papastavrou et~al.(2013a)Papastavrou, Steinmann and
		Kuhl}]{Papastavrou}
	\bibinfo{author}{Papastavrou, A.}, \bibinfo{author}{Steinmann, P.},
	\bibinfo{author}{Kuhl, E.}, \bibinfo{year}{2013}a.
	\newblock \bibinfo{title}{On the mechanics of continua with boundary energies
		and growing surfaces}.
	\newblock \bibinfo{journal}{Journal of the Mechanics and Physics of Solids}
	\bibinfo{volume}{61}, \bibinfo{pages}{1446--1463}.
	\newblock \DOIprefix\doi{10.1016/j.jmps.2013.01.007}.
	%Type = Article
	\bibitem[{Papastavrou et~al.(2013b)Papastavrou, Steinmann and
		Kuhl}]{papastavrou2013mechanics}
	\bibinfo{author}{Papastavrou, A.}, \bibinfo{author}{Steinmann, P.},
	\bibinfo{author}{Kuhl, E.}, \bibinfo{year}{2013}b.
	\newblock \bibinfo{title}{On the mechanics of continua with boundary energies
		and growing surfaces}.
	\newblock \bibinfo{journal}{Journal of the Mechanics and Physics of Solids}
	\bibinfo{volume}{61}, \bibinfo{pages}{1446--1463}.
	%Type = Article
	\bibitem[{Robertson et~al.(2018)Robertson, Yourdkhani, Centellas, Aw, Ivanoff,
		Goli, Lloyd, Dean, Sottos, Geubelle et~al.}]{robertson2018rapid}
	\bibinfo{author}{Robertson, I.D.}, \bibinfo{author}{Yourdkhani, M.},
	\bibinfo{author}{Centellas, P.J.}, \bibinfo{author}{Aw, J.E.},
	\bibinfo{author}{Ivanoff, D.G.}, \bibinfo{author}{Goli, E.},
	\bibinfo{author}{Lloyd, E.M.}, \bibinfo{author}{Dean, L.M.},
	\bibinfo{author}{Sottos, N.R.}, \bibinfo{author}{Geubelle, P.H.}, et~al.,
	\bibinfo{year}{2018}.
	\newblock \bibinfo{title}{Rapid energy-efficient manufacturing of polymers and
		composites via frontal polymerization}.
	\newblock \bibinfo{journal}{Nature} \bibinfo{volume}{557},
	\bibinfo{pages}{223}.
	%Type = Article
	\bibitem[{Sadhu and Dhar(2012)}]{sadhu2012modelling}
	\bibinfo{author}{Sadhu, T.}, \bibinfo{author}{Dhar, D.}, \bibinfo{year}{2012}.
	\newblock \bibinfo{title}{Modelling proportionate growth}.
	\newblock \bibinfo{journal}{Curr. Sci} \bibinfo{volume}{103}.
	%Type = Article
	\bibitem[{Savin et~al.(2011)Savin, Kurpios, Shyer, Florescu, Liang, Mahadevan
		and Tabin}]{savin2011growth}
	\bibinfo{author}{Savin, T.}, \bibinfo{author}{Kurpios, N.A.},
	\bibinfo{author}{Shyer, A.E.}, \bibinfo{author}{Florescu, P.},
	\bibinfo{author}{Liang, H.}, \bibinfo{author}{Mahadevan, L.},
	\bibinfo{author}{Tabin, C.J.}, \bibinfo{year}{2011}.
	\newblock \bibinfo{title}{On the growth and form of the gut}.
	\newblock \bibinfo{journal}{Nature} \bibinfo{volume}{476},
	\bibinfo{pages}{57--62}.
	%Type = Article
	\bibitem[{Skalak et~al.(1982)Skalak, Dasgupta, Moss, Otten, Dullemeijer and
		Vilmann}]{Skalak82}
	\bibinfo{author}{Skalak, R.}, \bibinfo{author}{Dasgupta, G.},
	\bibinfo{author}{Moss, M.}, \bibinfo{author}{Otten, E.},
	\bibinfo{author}{Dullemeijer, P.}, \bibinfo{author}{Vilmann, H.},
	\bibinfo{year}{1982}.
	\newblock \bibinfo{title}{Analytical description of growth}.
	\newblock \bibinfo{journal}{Journal of Theoretical Biology}
	\bibinfo{volume}{94}, \bibinfo{pages}{555--577}.
	\newblock \DOIprefix\doi{10.1016/0022-5193(82)90301-0}.
	%Type = Article
	\bibitem[{Skalak et~al.(1997)Skalak, Farrow and Hoger}]{Skalak97}
	\bibinfo{author}{Skalak, R.}, \bibinfo{author}{Farrow, D.},
	\bibinfo{author}{Hoger, A.}, \bibinfo{year}{1997}.
	\newblock \bibinfo{title}{Kinematics of surface growth}.
	\newblock \bibinfo{journal}{Journal of mathematical biology}
	\bibinfo{volume}{35}, \bibinfo{pages}{869--907}.
	\newblock \DOIprefix\doi{10.1007/s002850050081}.
	%Type = Article
	\bibitem[{Slaughter et~al.(2009)Slaughter, Khurshid, Fisher, Khademhosseini and
		Peppas}]{slaughter}
	\bibinfo{author}{Slaughter, B.V.}, \bibinfo{author}{Khurshid, S.S.},
	\bibinfo{author}{Fisher, O.Z.}, \bibinfo{author}{Khademhosseini, A.},
	\bibinfo{author}{Peppas, N.A.}, \bibinfo{year}{2009}.
	\newblock \bibinfo{title}{Hydrogels in regenerative medicine}.
	\newblock \bibinfo{journal}{Advanced materials} \bibinfo{volume}{21},
	\bibinfo{pages}{3307--3329}.
	%Type = Article
	\bibitem[{Sorg et~al.(2017)Sorg, Tilkorn, Hager, Hauser and
		Mirastschijski}]{sorg2017skin}
	\bibinfo{author}{Sorg, H.}, \bibinfo{author}{Tilkorn, D.J.},
	\bibinfo{author}{Hager, S.}, \bibinfo{author}{Hauser, J.},
	\bibinfo{author}{Mirastschijski, U.}, \bibinfo{year}{2017}.
	\newblock \bibinfo{title}{Skin wound healing: an update on the current
		knowledge and concepts}.
	\newblock \bibinfo{journal}{European Surgical Research} \bibinfo{volume}{58},
	\bibinfo{pages}{81--94}.
	%Type = Article
	\bibitem[{Sozio and Yavari(2017)}]{sozio2017}
	\bibinfo{author}{Sozio, F.}, \bibinfo{author}{Yavari, A.},
	\bibinfo{year}{2017}.
	\newblock \bibinfo{title}{Nonlinear mechanics of surface growth for cylindrical
		and spherical elastic bodies}.
	\newblock \bibinfo{journal}{Journal of the Mechanics and Physics of Solids}
	\bibinfo{volume}{98}, \bibinfo{pages}{12--48}.
	\newblock \DOIprefix\doi{10.1016/j.jmps.2016.08.012}.
	%Type = Article
	\bibitem[{Sozio and Yavari(2019)}]{sozio2019nonlinear}
	\bibinfo{author}{Sozio, F.}, \bibinfo{author}{Yavari, A.},
	\bibinfo{year}{2019}.
	\newblock \bibinfo{title}{Nonlinear mechanics of accretion}.
	\newblock \bibinfo{journal}{Journal of Nonlinear Science} \bibinfo{volume}{29},
	\bibinfo{pages}{1813--1863}.
	\newblock \DOIprefix\doi{10.1007/s00332-019-09531-w}.
	%Type = Article
	\bibitem[{Stracuzzi et~al.(2018)Stracuzzi, Mazza and Ehret}]{stracuzzi2018}
	\bibinfo{author}{Stracuzzi, A.}, \bibinfo{author}{Mazza, E.},
	\bibinfo{author}{Ehret, A.E.}, \bibinfo{year}{2018}.
	\newblock \bibinfo{title}{Chemomechanical models for soft tissues based on the
		reconciliation of porous media and swelling polymer theories}.
	\newblock \bibinfo{journal}{Journal of Applied Mathematics and Mechanics}
	\bibinfo{volume}{98}, \bibinfo{pages}{2135--2154}.
	\newblock \DOIprefix\doi{10.1002/zamm.201700344}.
	%Type = Article
	\bibitem[{Swain and Gupta(2018)}]{swain2018biological}
	\bibinfo{author}{Swain, D.}, \bibinfo{author}{Gupta, A.}, \bibinfo{year}{2018}.
	\newblock \bibinfo{title}{Biological growth in bodies with incoherent
		interfaces}.
	\newblock \bibinfo{journal}{Proceedings of the Royal Society A: Mathematical,
		Physical and Engineering Sciences} \bibinfo{volume}{474},
	\bibinfo{pages}{20170716}.
	%Type = Article
	\bibitem[{Tallinen et~al.(2016)Tallinen, Chung, Rousseau, Girard, Lef{\`e}vre
		and Mahadevan}]{tallinen2016growth}
	\bibinfo{author}{Tallinen, T.}, \bibinfo{author}{Chung, J.Y.},
	\bibinfo{author}{Rousseau, F.}, \bibinfo{author}{Girard, N.},
	\bibinfo{author}{Lef{\`e}vre, J.}, \bibinfo{author}{Mahadevan, L.},
	\bibinfo{year}{2016}.
	\newblock \bibinfo{title}{On the growth and form of cortical convolutions}.
	\newblock \bibinfo{journal}{Nature Physics} \bibinfo{volume}{12},
	\bibinfo{pages}{588--593}.
	%Type = Book
	\bibitem[{Thompson(1917)}]{Thompson}
	\bibinfo{author}{Thompson, D.W.}, \bibinfo{year}{1917}.
	\newblock \bibinfo{title}{On growth and form}.
	\newblock \bibinfo{publisher}{University Press}, \bibinfo{address}{Cambridge
		[Eng.]}.
	%Type = Article
	\bibitem[{Tomassetti et~al.(2016)Tomassetti, Cohen and Abeyaratne}]{TCA}
	\bibinfo{author}{Tomassetti, G.}, \bibinfo{author}{Cohen, T.},
	\bibinfo{author}{Abeyaratne, R.}, \bibinfo{year}{2016}.
	\newblock \bibinfo{title}{Steady accretion of an elastic body on a hard
		spherical surface and the notion of a four-dimensional reference space}.
	\newblock \bibinfo{journal}{Journal of the Mechanics and Physics of Solids}
	\bibinfo{volume}{96}, \bibinfo{pages}{333--352}.
	\newblock \DOIprefix\doi{10.1016/j.jmps.2016.05.015}.
	%Type = Article
	\bibitem[{Truster and Masud(2017)}]{truster2017unified}
	\bibinfo{author}{Truster, T.J.}, \bibinfo{author}{Masud, A.},
	\bibinfo{year}{2017}.
	\newblock \bibinfo{title}{A unified mixture formulation for density and
		volumetric growth of multi-constituent solids in tissue engineering}.
	\newblock \bibinfo{journal}{Computer Methods in Applied Mechanics and
		Engineering} \bibinfo{volume}{314}, \bibinfo{pages}{222--268}.
	\newblock \DOIprefix\doi{10.1016/j.cma.2016.09.023}.
	%Type = Article
	\bibitem[{Turing(1952)}]{turing1952chemical}
	\bibinfo{author}{Turing, A.M.}, \bibinfo{year}{1952}.
	\newblock \bibinfo{title}{The chemical basis of morphogenesis}.
	\newblock \bibinfo{journal}{Philosophical Transactions of the Royal Society of
		London. Series B, Biological Sciences} \bibinfo{volume}{237},
	\bibinfo{pages}{37--72}.
	\newblock \DOIprefix\doi{10.1098/rstb.1952.0012}.
	%Type = Article
	\bibitem[{Wang et~al.(2020)Wang, Zhang, Zheng and Ye}]{wang2020high}
	\bibinfo{author}{Wang, J.}, \bibinfo{author}{Zhang, H.},
	\bibinfo{author}{Zheng, Y.}, \bibinfo{author}{Ye, H.}, \bibinfo{year}{2020}.
	\newblock \bibinfo{title}{High-order nurbs elements based isogeometric
		formulation for swellable soft materials}.
	\newblock \bibinfo{journal}{Computer Methods in Applied Mechanics and
		Engineering} \bibinfo{volume}{363}, \bibinfo{pages}{112901}.
	\newblock \DOIprefix\doi{10.1016/j.cma.2020.112901}.
	%Type = Article
	\bibitem[{Zurlo and Truskinovsky(2017)}]{zurlo2017printing}
	\bibinfo{author}{Zurlo, G.}, \bibinfo{author}{Truskinovsky, L.},
	\bibinfo{year}{2017}.
	\newblock \bibinfo{title}{Printing non-euclidean solids}.
	\newblock \bibinfo{journal}{Phys. Rev. Lett.} \bibinfo{volume}{119},
	\bibinfo{pages}{048001}.
	\newblock \DOIprefix\doi{10.1103/PhysRevLett.119.048001}.
	%Type = Article
	\bibitem[{Zurlo and Truskinovsky(2018)}]{zurlo2018inelastic}
	\bibinfo{author}{Zurlo, G.}, \bibinfo{author}{Truskinovsky, L.},
	\bibinfo{year}{2018}.
	\newblock \bibinfo{title}{Inelastic surface growth}.
	\newblock \bibinfo{journal}{Mechanics Research Communications}
	\bibinfo{volume}{93}, \bibinfo{pages}{174--179}.
	\newblock \DOIprefix\doi{10.1016/j.mechrescom.2018.01.007}.
	
\end{thebibliography}

\end{document}